\documentclass[a4paper]{mn2e}

\usepackage{times}
\usepackage{epsfig}
\usepackage{subfigure}
\usepackage{astrojournals}

\title{An alternative to common envelope evolution}

\author[M.\,E.\,Beer, L.\,M.\,Dray, A.\,R.\,King \& G.\,A.\,Wynn]{Martin
E. Beer\thanks{E-mail: martin.beer@astro.le.ac.uk}, Lynnette M. Dray, Andrew
R. King and Graham A. Wynn\\ Department of Physics and Astronomy, University
of Leicester, Leicester, LE1 7RH, United Kingdom}

\def\today{\number\year \ \ifcase\month\or
  January\or February\or March\or April\or May\or June\or
  July\or August\or September\or October\or November\or December
 \fi \ \number\day }
\date{Accepted 2006 December 1. Received 2006 November 14; in original form 2006 May 8}

\volume{000}

\pagerange{\pageref{firstpage}--\pageref{lastpage}} \pubyear{2006}

\begin{document}

\label{firstpage}

\maketitle

\begin{abstract}
We investigate the evolution of interacting binaries where the donor
star is a low--mass giant more massive than its companion. It is usual
to assume that such systems undergo common--envelope (CE) evolution,
where the orbital energy is used to eject the donor envelope, thus
producing a closer binary or a merger. We suggest instead that because
mass transfer is super--Eddington even for non--compact companions, a
wide range of systems avoid this type of CE phase. The accretion
energy released in the rapid mass transfer phase unbinds a significant
fraction of the giant's envelope, reducing the tendency to dynamical
instability and merging. We show that our physical picture accounts
for the success of empirical parametrizations of the outcomes of
assumed CE phases.
\end{abstract}

\begin{keywords}
binaries: close -- stars: evolution
\end{keywords}

\section{Introduction}
Binary systems tend to shrink when they transfer mass from the more
massive to the less massive star. The consequent contraction of the
donor's Roche lobe leads to high transfer rates, proceeding on a
thermal or near--dynamical timescale depending on whether this star is
predominantly radiative or convective. There is evidently a tendency
for such systems to enter common envelope (CE) evolution, where the
donor star engulfs the system \nocite{pac76,ost76,w84}({Paczy{\'n}ski} 1976; {Ostriker} 1976; {Webbink} 1984). The archetypal
case arises when the donor star fills its Roche lobe as a giant. The
strong dissipation caused by the accretor moving within the common
envelope may then produce two effects: (a) the binary orbit shrinks,
releasing orbital binding energy, and (b) this energy release may
unbind the envelope entirely, stopping the process before the accretor
merges with the core of the donor. Features (a) and (b) are highly
desirable in explaining the formation of cataclysmic variables (CVs),
where a low--mass main sequence star orbits a white dwarf at a
separation considerably smaller than the radius of the white dwarf's
giant progenitor, and indeed for tight binary formation in general.

Despite this early promise, several decades of strenuous effort have
failed to provide convincing evidence that the process works as
originally hoped \nocite{ts00}(see {Taam} \&  {Sandquist} 2000, and references
  therein). Technically the problem is difficult, requiring 
full 3D hydrodynamics and a careful treatment of the dissipation
processes. However there appear to be problems at a more basic level,
which are manifest when we consider the most popular parametrization
of CE \nocite{w84}(e.g. {Webbink} 1984). This compares the envelope
binding energy with the change in orbital energy as 
\begin{equation} 
{GM_{\rm g}M_{\rm e}\over \lambda R_{\rm g}} = \alpha\left({GM_{\rm
    c}M_2\over 2a_{\rm f}} - {GM_gM_2\over 2a_{\rm
    i}}\right). \label{en}
\end{equation} 
Here $M_{\rm g}$ is the giant mass, $R_{\rm g}$ its radius, $M_{\rm
c}$ its core mass, $M_{\rm e}$ its envelope mass, $M_2$ is the
accretor mass (assumed fixed during CE evolution) and $a_{\rm
i},~a_{\rm f}$ are the initial and final binary separations. Stellar
structure calculations specify the dimensionless parameter $\lambda<
1$, which is often taken as a constant, e.g. $\lambda =0.5$, and
$\alpha < 1$ is a dimensionless efficiency parameter. Clearly only the
combination $\alpha\lambda$ is important. The problem which emerges is
that in many cases the post--CE orbital separation $a_{\rm f}$ is too
large, i.e. there was too little orbital energy release to unbind the
giant envelope. Formally this problem appears in the requirement
$\alpha\lambda >1$ (values $\alpha\lambda \sim 6$ are not
unknown). \nocite{nt05}{Nelemans} \& {Tout} (2005) show that these
problems are still worse when considering the formation of double
white dwarf binaries, where it is possible to reconstruct the state of
the progenitor binaries. The required values of $\alpha\lambda$ are
actually negative in most cases. All these difficulties point to the
need for an extra source of energy to unbind the giant envelope.

A clue to this energy source comes from an observed case which CE
evolution clearly cannot produce. The low--mass X--ray binary Cyg X--2
consists of a neutron star accreting from a low--mass donor ($\simeq$
0.5\,--\,0.7\,${\rm M_{\sun}}$) in a 9.8\,d orbit \nocite{kr99}(see
{King} \& {Ritter} 1999, and references therein). The donor has radius
$\simeq 7\,{\rm R_{\sun}}$, and its effective temperature implies
luminosity $\simeq 150\,{\rm L_{\sun}}$. Evidently this star was
considerably more massive in the recent past and we are now seeing its
luminous inner layers before they cool. But the current binary period
of 9.8~d makes the binary so wide that far too little orbital energy
was available to unbind the envelope, regardless of how wide the
pre--CE separation $a_{\rm i}$ was. However in this case the required
extra energy source is clear. Assuming that mass transfer began when
the donor was the more massive star and still predominantly radiative,
mass flowed towards the neutron star on a thermal timescale $\sim
10^6$\,yr, giving a transfer rate $\ga 10^{-6}\,{\rm
M_{\sun}}$\,yr$^{-1}$. This is clearly highly
super--Eddington. \nocite{kr99}{King} \& {Ritter} (1999) were able to
suggest a plausible evolutionary sequence leading to the current state
of Cyg X--2 by assuming that the neutron star accreted only at the
Eddington rate, the remaining transferred mass being ejected from the
system with the specific angular momentum of the neutron
star. \nocite{pr00}{Podsiadlowski} \& {Rappaport} (2000) similarly
found a believable evolutionary history for the system starting from a
somewhat less evolved donor.

The crucial step in understanding Cyg X--2 is to assume that the
accretion energy of some of the transferred matter can eject the
remainder, without the system entering CE evolution. Although earlier
papers considered the possibility of using accretion energy in this
way \nocite{hw99}(cf {Han} \&  {Webbink} 1999) they always concluded
that the radiatively driven outflow would be dense and extended enough
to act as a surrogate envelope for the accretor, and lead back to CE
evolution.  However \nocite{kb99}{King} \&  {Begelman} (1999) argued
that CE evolution would only occur if a sufficiently dense region of
the gas flow on to the accretor overfilled the latter's Roche
lobe. They identified the boundary of this dense region as
the trapping radius
\begin{equation}
  R_{\rm trap} = \frac{\left|\dot{M}_1\right|} {\dot{M}_{\rm edd}} R_2
  \,, \label{etrap} 
\end{equation} 
where the accretion flow first reaches its local Eddington limit
($\dot{M}_{\rm edd}$ is the Eddington rate at the accretor radius
$R_2$, see below) overfilled the accretor's Roche lobe. The trapping
radius is effectively identical to the spherization radius defined for
disc accretion by \nocite{ss73}{Shakura} \& {Sunyaev} (1973), who show
that disc accretion at super-Eddington rates becomes spherical inside
this radius and is a tenuous wind outside it, making it a reasonable
estimate for the effective size of any dense envelope (see also
section~\ref{sect_method}).

On this basis \nocite{kb99}{King} \& {Begelman} (1999) showed that
systems where a neutron star or black hole accretes from a more
massive star would probably always avoid CE evolution provided that
the donor was predominantly radiative, i.e. that mass transfer
proceeded on a thermal timescale, as in Cyg X--2. This allows values
as high as $10^{-3}$ M$_{\sun}$\,yr$^{-1}$, and SS433 is an example of
a system in this state \nocite{ktb00,bkp06}({King}, {Taam} \&
{Begelman} 2000; {Begelman}, {King} \& {Pringle}
2006). \nocite{kb99}{King} \& {Begelman} (1999) also suggested that
such avoidance of CE evolution might apply even to donors which were
largely convective, but did not attempt quantitative estimates.

In this paper we extend the treatment of \nocite{kb99}{King} \&
{Begelman} (1999) to such cases. We note that the mass transfer rates
probably exceed the Eddington accretion rate
\begin{equation}
  \dot{M}_{\rm edd} = \frac{4 \pi R_2 m_{\rm p} c} {\sigma_{\rm T}}
\sim 10^{-3}{R_2\over {\rm R_{\sun}}}{\rm M_{\sun}}\, {\rm yr}^{-1}
\label{edd}
\end{equation}
for gravitational energy release on a main sequence star (here $R_2$ is the
radius of the accretor, $m_{\rm p}$ is the mass of a proton, $c$ the speed of
light and $\sigma_{\rm T}$ the Thompson scattering cross-section). Thus it may
be possible to avoid CE evolution even in cases where a red giant overflows on
to a main sequence star, as required to form cataclysmic variables.

Specifically we investigate here the evolution of systems in which
mass transfer occurs from giants on to less massive companions. We
find that the trapping radius of the companion (where mass accretion
reaches the local Eddington limit) is within its Roche lobe, so CE
evolution may be avoided. Our treatment suggests that a comparison of
the fractional change of mass with angular momentum offers a more apt
parametrization of the change of orbital separation than the energy
formalism (equation~\ref{en}). We note that \nocite{nt05}{Nelemans} \&
{Tout} (2005) have proposed such a formalism (but without an explicit
physical reason for the mass loss) which empirically describes the
formation of sub--dwarf B star/white--dwarf and white--dwarf/M--dwarf
binaries. We shall show that our physical treatment gives similar
results, thus offering a possible explanation for its success in this
case.

\section{Evolutionary Method} \label{sect_method}
We consider donor stars in initial evolutionary states ranging from
the start of the Hertzsprung gap to the tip of the asymptotic giant
branch.

We define the mass ratio $q = M_1/M_2$ (normally $>1$) with $M_1$ the
donor mass and $M_2$ the accretor mass. The Roche lobe radius ($R_{\rm
L}$) as a function of orbital separation ($a$) for the donor is given
by \nocite{egg83}{Eggleton} (1983)
\begin{equation}
  \frac{R_{\rm L}}{a} = \frac {0.49 q^{2/3}} {0.6 q^{2/3} + \ln (1 +
    q^{1/3}) } \,. \label{eqeggrl}
\end{equation}

For given masses we can find the radius of the donor when it fills its
Roche lobe at a given initial period ($P_{\rm i}$). Using the Eggleton
stellar evolution code \nocite{egg71,pet98}({Eggleton} 1971; {Pols}
et~al. 1998, and references therein) we then find the corresponding
core mass ($M_{\rm c}$) of the donor at this radius. We assume that
mass transfer is so rapid that the core mass does not significantly
increase during the mass transfer phase.

The instantaneous mass transfer rate is given by tracking the
overfilling of the donor's Roche lobe as given by
\nocite{ps72}{Paczy{\'n}ski} \& {Sienkiewicz} (1972). Substituting
their equation (4) into their equation (A26) we find
\begin{equation}
  -\dot{M}_1 = \frac{S_1 E(m) W(\mu)} {2 P_{\rm orb}} \left ( \frac{R_1}{a}
  \right )^{-\frac{3}{2}} \frac{(M_1+M_2)^{\frac{3}{2}}} {\sqrt{M_1}} \left (
  \frac {\Delta R_1}{R_1} \right)^3 \label{eqmdot}
\end{equation}
where $S_1$ is 0.215; $m$ is the fractional core-mass of the donor
($M_{\rm{c}}/M_1$); $E(m)$ is a function tabulated by
\nocite{ps72}{Paczy{\'n}ski} \& {Sienkiewicz} (1972); $\mu$ is
$M_1/(M_1+M_2)$; $\Delta R_1$ is the difference in stellar radius and
Roche-lobe radius ($R_{\rm{L}}$) of the donor and $W(\mu)$ is given by
\begin{equation}
  W(\mu) = \frac{\sqrt{\mu}\sqrt{1-\mu}} {\left (
  \sqrt{\mu}+\sqrt{1+\mu} \, \right )^4} \left ( \frac {\mu} {R_{\rm{L}}}
  \right )^3 ~.
\end{equation}

We let a fraction $\psi$ of the transferred matter be blown away
\nocite{bh91}(cf {Bhattacharya} \& {van den Heuvel} 1991, who call
this fraction $\beta$). An expression for $\psi$ which accounts for
the proportion of matter which must be accreted to power the outflow
is given by \nocite{hw99}{Han} \& {Webbink} (1999) (who call this
quantity $1-\beta$)
\begin{eqnarray}
  \psi = \left \{
  \begin{array}{ll}
    1 - \frac{\hbox{\rule[-3pt]{0pt}{12pt}$L_{\rm edd}$}}
  {\hbox{\rule[-3pt]{0pt}{12pt}$\left|\phi_{\rm R2}\right| \,
  \left|\dot{M}_1\right|$}} -
  \frac{\hbox{\rule[-3pt]{0pt}{12pt}$\phi_{\rm L1}$}}
  {\hbox{\rule[-3pt]{0pt}{12pt}$\phi_{\rm R2}$}} & \rm{ if ~}
  \left|\dot{M}_1 \right|
        \ge \dot{M}_{\rm edd} \\
  0 & \rm{ if ~} \left|\dot{M}_1\right| < \dot{M}_{\rm edd} 
  \end{array} \right . ~,
  \label{eqbeta}
\end{eqnarray}
where $L_{\rm edd}$ is the Eddington luminosity 
\begin{equation}
  L_{\rm edd} = \frac{4\pi G M_2 m_{\rm p} c}{\sigma_{\rm T}} ~,
\end{equation}
and $\phi_{\rm L1}$ and $\phi_{\rm R2}$ are the potentials at the inner
Lagrangian point and the surface of the accretor respectively and are
given by
\begin{equation}
  \phi_{\rm{L1}} = -\frac{GM_1}{a-b} -\frac{GM_2}{b}
  -\frac{G(M_1+M_2)} {2a^3} (\mu a -b)^2 ~,
\end{equation}
\begin{equation}
    \phi_{\rm{R2}} = -\frac{GM_1}{a} -\frac{GM_2}{R_2} -
  \frac{G(M_1+M_2)} {2a^3} \left [ \frac{2}{3}R_2^2 + (\mu a)^2 \right
  ] ~,
\end{equation}
where $b$ is the distance from the accretor to the inner Lagrangian
point and is given by \nocite{w76}{Warner} (1976)
\begin{equation}
\frac{b}{a} = 0.5 - 0.227 \log q ~,
\end{equation}
if ~$0.1 \le q \le 10$ or \nocite{k59}{Kopal} (1959) if ~$q > 10$
\begin{equation}
  \frac{b}{a} = w - \frac{w^2}{3} - \frac{w^3}{9} ~,
\end{equation}
where
\begin{equation}
  w^3 = \frac{1}{3(1+q)} ~.
\end{equation}
If $R_2 \ll a$ equation~(\ref{eqbeta}) becomes for
$\left|\dot{M_1}\right| \ge \dot{M}_{\rm edd}$
\begin{equation}
  \psi = 1 - \frac{\dot{M}_{\rm edd}} {\left|\dot{M_1}\right|} ~.
\end{equation}
This is equivalent to assuming that the transferred matter cannot be
accreted at a rate faster than the Eddington rate
(equation~\ref{edd}), and that all transferred matter is accreted for
sub-Eddington transfer rates. We consider below how much angular momentum 
this mass carries off.

The trapping radius of the accretor, equation~(\ref{etrap}), 
depends only on the mass transfer rate from the donor. During the
evolution we compare the trapping radius with the Roche radius of the accretor
found using equation~(\ref{eqeggrl}) with the mass ratio now inverted. If the
trapping radius is greater than the Roche lobe we assume that the system
enters CE evolution and do not follow its evolution. In the rest of the paper
we show only systems where the trapping radius never overfills the Roche lobe.

\begin{figure}                                                              
\begin{center}                                                         
\epsfig{file=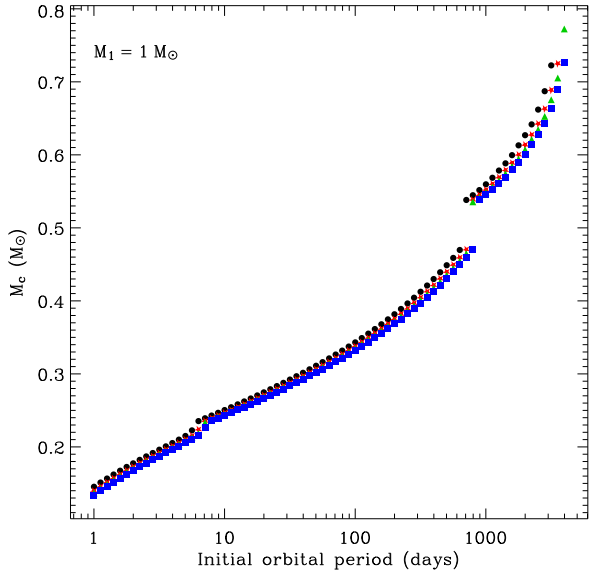,height=8cm}                               
\end{center}                                                          
\caption{The initial periods at which a 1\,M$_{\sun}$ donor fills         
  its Roche lobe                                                        
  and the corresponding donor core masses for various accretors:             
  0.2\,M$_{\sun}$ main-sequence stars (black circles), white dwarfs          
  (red stars), 0.6\,M$_{\sun}$ main-sequence stars (green triangles)        
  and 1\,M$_{\sun}$ main-sequence stars (blue
  squares).}\label{figperiodinit}   
\end{figure}                                                              
  
\begin{figure}                                                            
\begin{center}                                                     
\epsfig{file=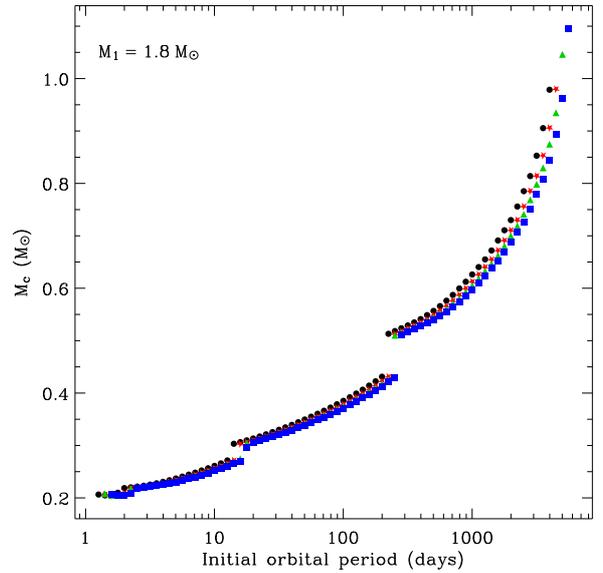,height=8cm}                               
\end{center}                                                 
\caption{The initial periods at which a 1.8\,M$_{\sun}$ donor fills   
  its Roche lobe                                                         
 and the corresponding donor core masses for various                       
  accretors. Symbols are the                                          
  same as in Fig.~\ref{figperiodinit}.}\label{figperiodinit18}    
\end{figure} 

\begin{figure*}                                                                         
\begin{center}                                                                          
\subfigure{\epsfig{file=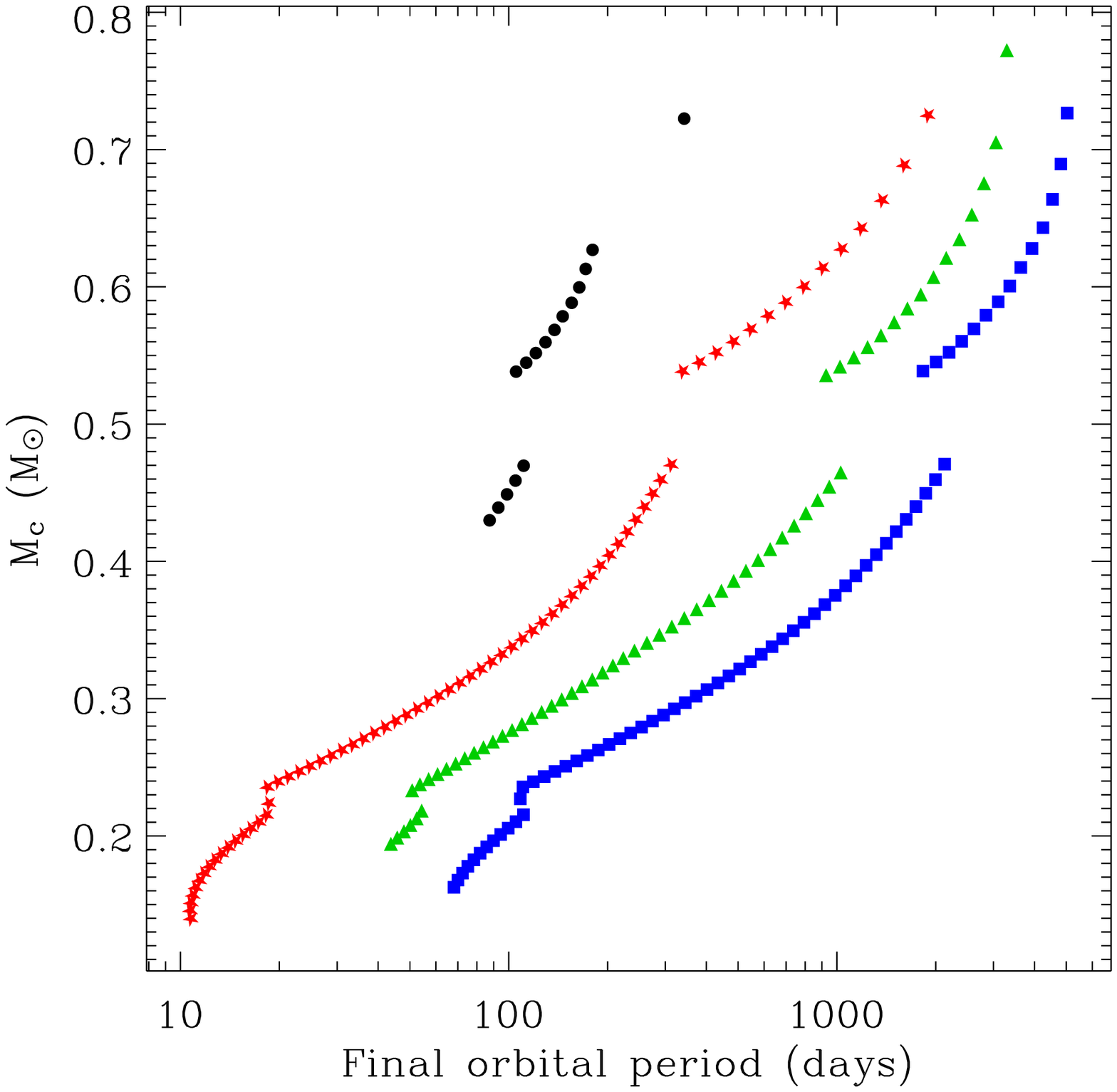,height=5.5cm}                                          
\label{figmp2}}                                                                         
\subfigure{\epsfig{file=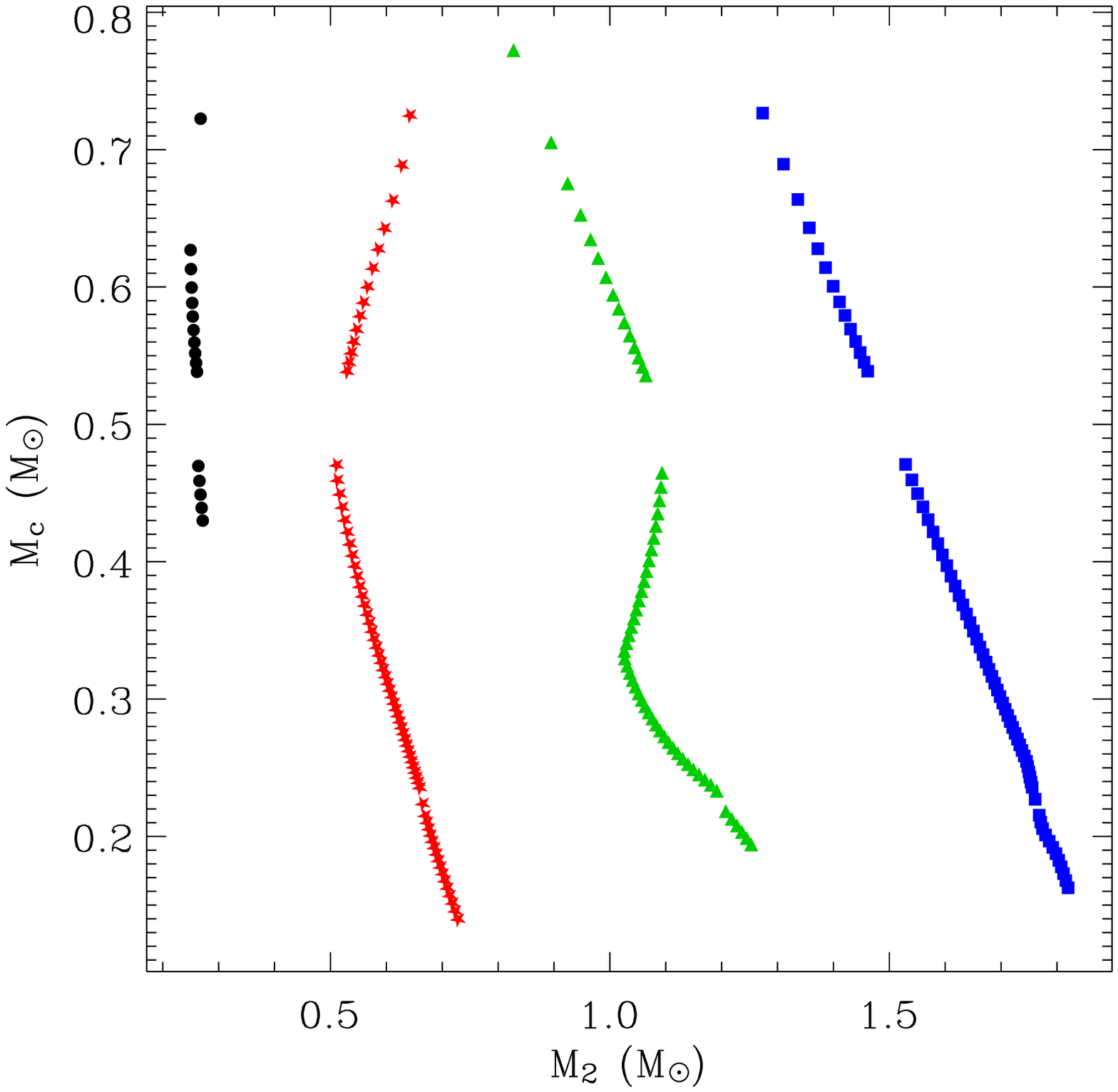,height=5.5cm}                                          
\label{figmm2}}                                                                         
\subfigure{\epsfig{file=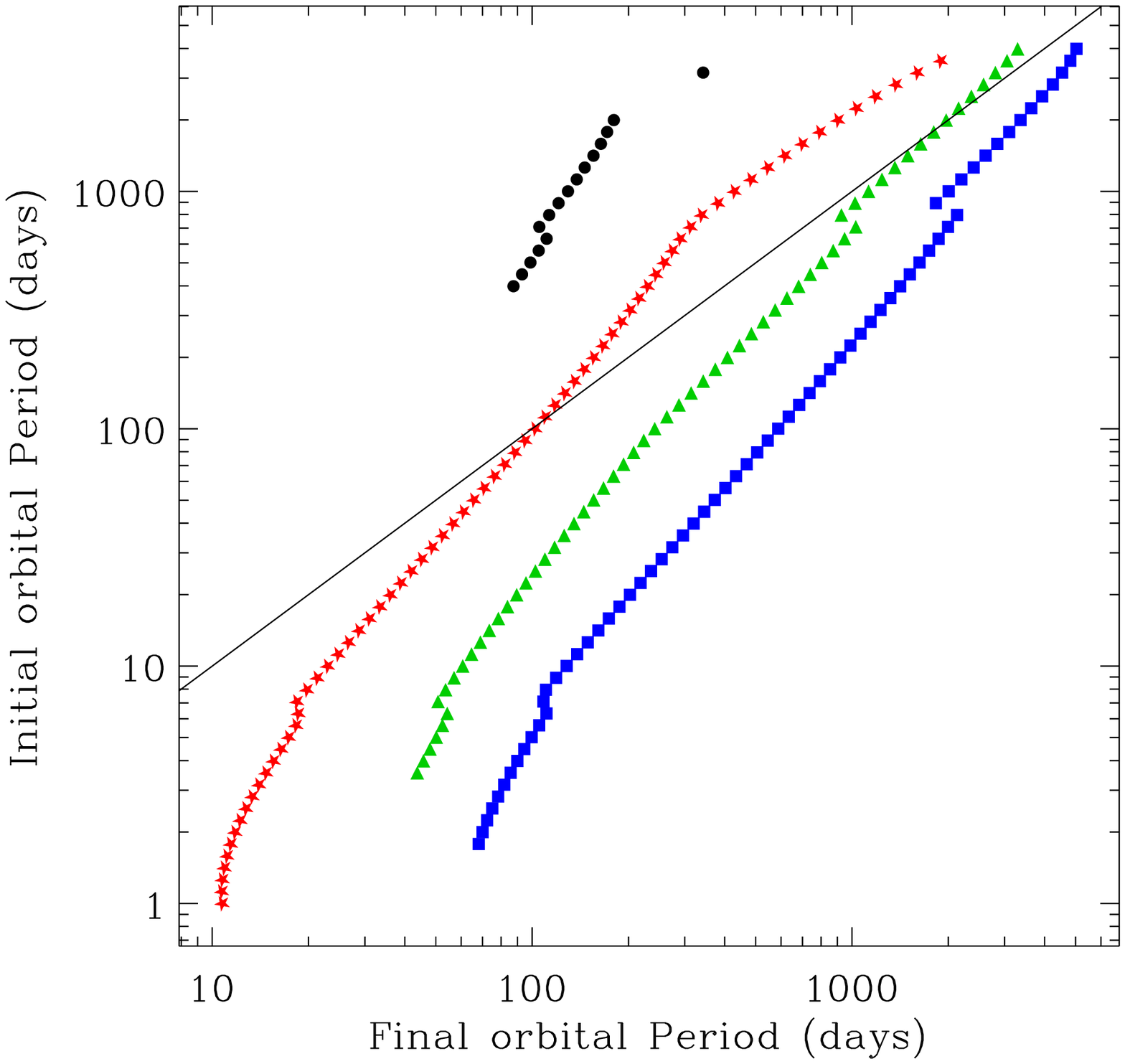,height=5.5cm}\label{figpp2}}                           
\vspace{-0.3cm}                                                                         
\caption{Final period and accretor mass versus core mass, and                           
  final versus initial period, for mass-transfer from a 1\,M$_{\sun}$                   
  donor. Symbols as in                                                                  
  Fig.~\ref{figperiodinit}. Diagonal lines indicate where the final                     
  and initial orbital periods are equal.}                                               
\label{figseq100yr}                                                                     
\end{center}                                                                            
\end{figure*}                                                                           
                                                                                        
\begin{figure*}                                                                         
\vspace{-0.7cm}                                                                         
\begin{center}                                                                          
\subfigure{\epsfig{file=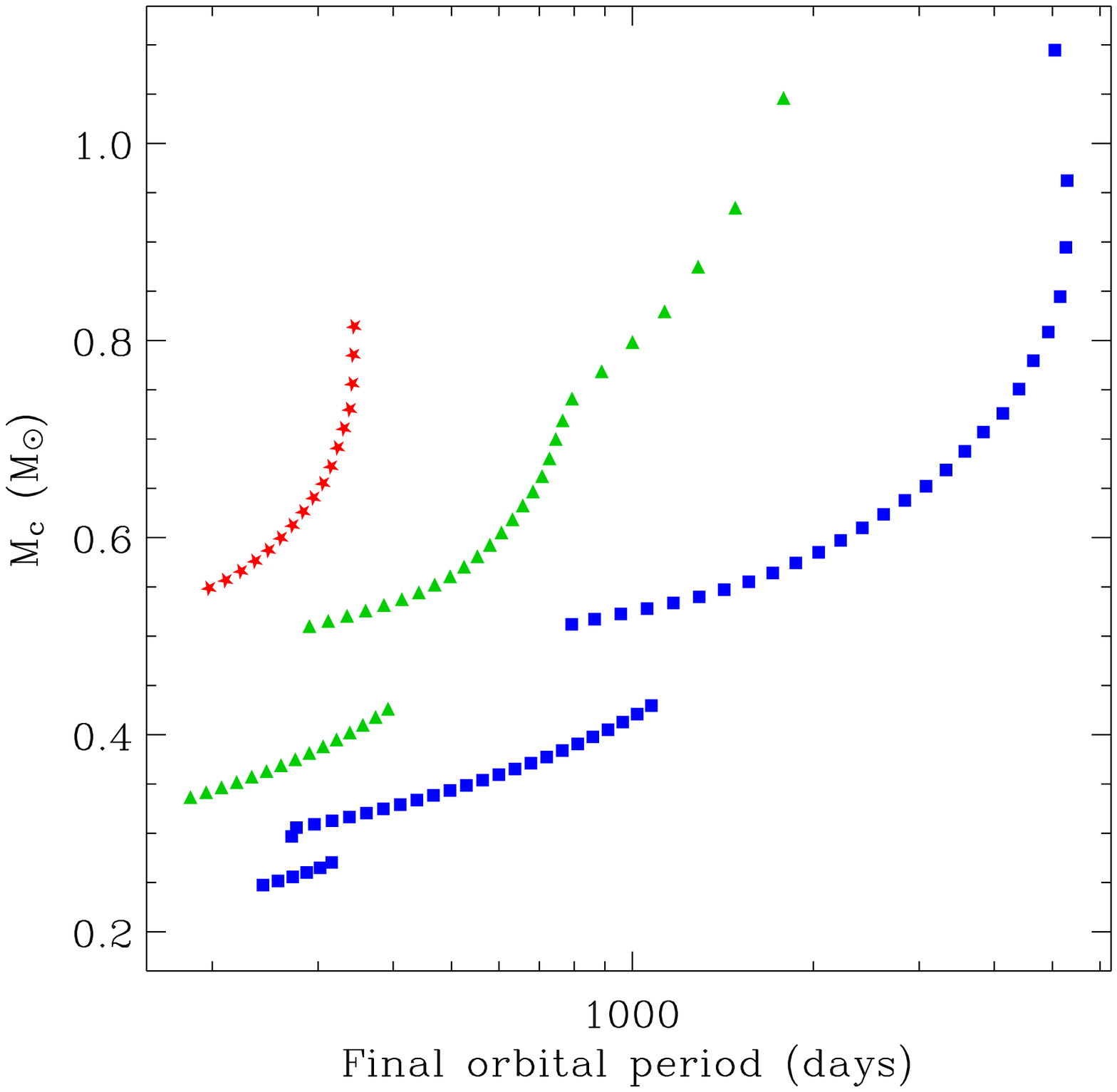,height=5.5cm}                                          
\label{fig18mp2}}                                                                       
\subfigure{\epsfig{file=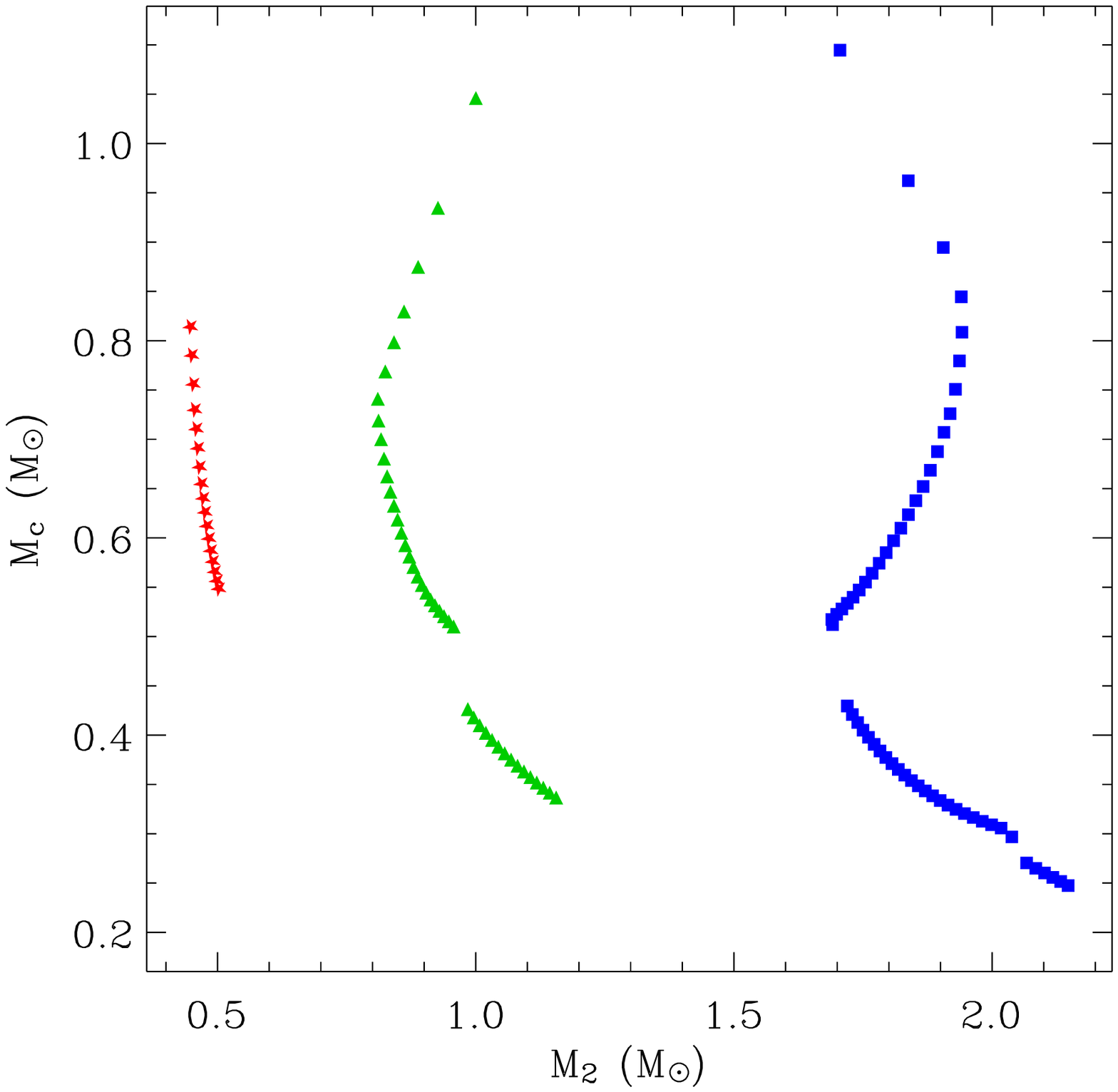,height=5.5cm}                                          
\label{fig18mm2}}                                                                       
\subfigure{\epsfig{file=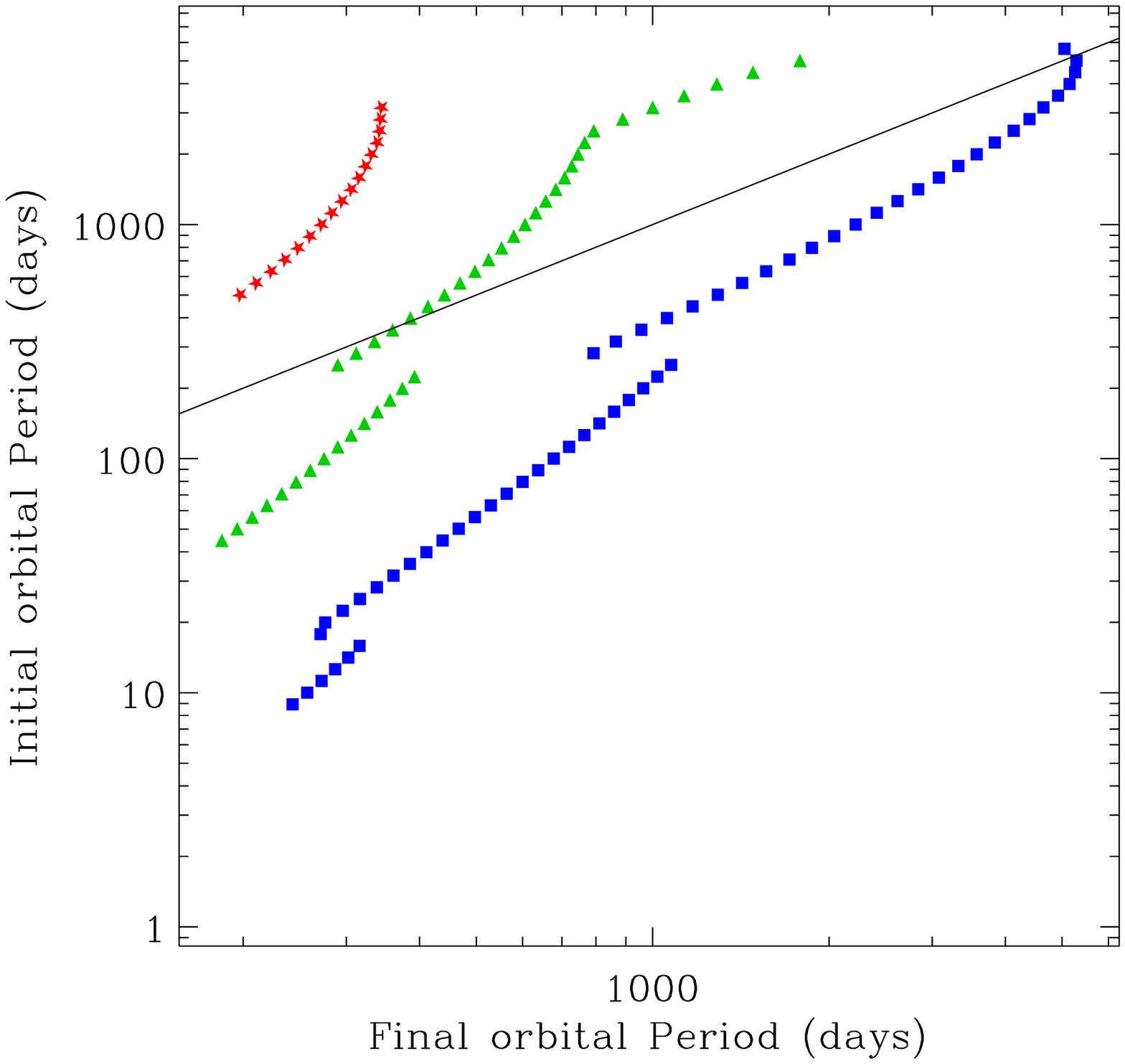,height=5.5cm}\label{fig18pp2}}                         
\vspace{-0.3cm}                                                                         
\caption{Final period and accretor mass versus core mass, and                           
  final versus initial period, for mass-transfer timescale from a                       
  1.8\,M$_{\sun}$ donor. Symbols as in                                                  
  Fig.~\ref{figperiodinit}. Diagonal lines indicate where the final                     
  and initial orbital periods are equal.}                                               
\label{figseq100yr1.8}                                                                  
\end{center}                                                                            
\end{figure*}  

The resultant change in orbital separation depends on the specific angular
momentum of the lost mass. If the transferred matter forms a disc and mass is
lost with circular symmetry from this disc, the appropriate value is the
specific angular momentum of the accretor. As the accretor is further from the
centre of mass this process shrinks the binary. However if the transferred
matter cannot circularize and form a disc, i.e. its circularization radius
\begin{equation}
R_{\rm circ} \simeq a (1+q) \left ( \frac{b}{a} \right) ^4
\label{eqrcirc} 
\end{equation}
(where $b$ is the distance of the accretor from the inner Lagrange
point $L_1$, cf \nocite{fkr02}{Frank}, {King} \& {Raine} 2002) is
smaller than the trapping radius $R_{\rm trap}$, the lost matter has
specific angular momentum closer to that of $L_1$. This follows
because, unless a disc forms, the gas stream from the donor
effectively conserves its specific angular momentum at a value close
to that of $L_1$ \nocite{kwf90}({King}, {Whitehurst} \& {Frank} 1990).
The donor thus loses matter with specific angular momentum slightly
less than its centre of mass, i.e. the donor slightly increases its
specific angular momentum. Since the binary mass also drops, this
tends to lead to expansion, as equation (20) of \nocite{tk98}{van
Teeseling} \& {King} (1998) shows.  This of course then makes $R_{\rm
circ}$ increase above $R_{\rm trap}$, creating a stable feedback so
that the binary subsequently evolves with $R_{\rm circ} \simeq R_{\rm
trap}$. This occurs particularly in systems with large mass ratios,
where most of the transferred matter is lost. These systems thus do
not merge unless the trapping radius exceeds the Roche radius of the
accretor and a CE phase occurs.


To demonstrate the trapping radius is the boundary of the dense region
we compute the density $\rho(R)$ in the outflow from the trapping
radius. This is essentially a spherical wind with constant velocity
equal to the escape velocity $v_{\rm esc}(R_{\rm trap} = (2GM/R_{\rm
trap})^{1/2}$ from the trapping radius. Thus
\begin{equation}
\rho(R) = {\dot M_{\rm out}\over 4\pi R^2v_{\rm esc}(R_{\rm trap})}.
\end{equation}
Now using (\ref{etrap}) gives
\begin{equation}
\rho = {\dot M_{\rm edd}\over 4\pi(GM_2)^{1/2}R_2^{3/2}}\biggl({\dot
  M_{\rm edd}\over \dot M_{\rm out}}\biggr)^{1/2}\biggl({R_{\rm
  trap}\over R}\biggr)^2.
\end{equation}
Here we use (\ref{edd}) to replace $\dot M_{\rm edd}$ in the first
factor on the rhs, giving
\begin{equation}
\rho = {cm_p\over \sigma_{\rm T}(GM_2R_2)^{1/2}}\biggl({\dot
  M_{\rm edd}\over \dot M_{\rm out}}\biggr)^{1/2}\biggl({R_{\rm
  trap}\over R}\biggr)^2
\label{tau}
\end{equation}
or 
\begin{equation}
\rho = {3.8\times 10^{-8}\over (m_2r_2)^{1/2}}\biggl({\dot
  M_{\rm edd}\over \dot M_{\rm out}}\biggr)^{1/2}\biggl({R_{\rm
  trap}\over R}\biggr)^2~{\rm g\ cm^{-3}}
\label{rho}
\end{equation}
where $m_2, r_2$ are $M_2, R_2$ in solar units. (Note that eqn
\ref{tau} reproduces the familiar result that the optical depth
through an outflow at the Eddington rate is of order unity at the
Schwarzschild radius, i.e. setting $\dot M_{\rm out} = \dot M_{\rm
edd}, R_2 = R_{\rm trap} = R_s = 2GM/c^2$ gives $\tau =
\rho\sigma_{\rm T}R_s/m_p \sim 1$.)

%
The density (\ref{rho}) implies a drag force $\sim \pi R_2^2\rho v^2$,
where $R_2, v$ are the companion's radius and orbital velocity, or
equivalently a drag torque
\begin{equation}
-\dot J \sim \pi R_2^2a\rho v^2
\end{equation}
where $a$ is the separation. The companion's angular momentum is
\begin{equation}
J \sim (4/3)\pi \rho_2f^3a^3(a^2/P)
\end{equation}
where $\rho_2$ is its mean density (writing its radius as $fa$). Using
this in $\dot J$ we get the drag timescale as
\begin{equation}
t_{\rm drag} \sim -{J\over \dot J} \sim {\rho_2 \over \rho}fP
\end{equation}
In other words, the drag is slow compared with the dynamical timescale $P$
provided that $\rho \ll \rho_2$. Since the companion is (at worst) a
MS star with $\rho_2 \sim 1$ we see from (\ref{rho}) that the drag is
small provided $a \gg R_{\rm trap}$ since $\dot M_{\rm out} \gg \dot
M_{\rm edd}$ by hypothesis.


\begin{figure}                                                                          
\begin{center}                                                                          
\epsfig{file=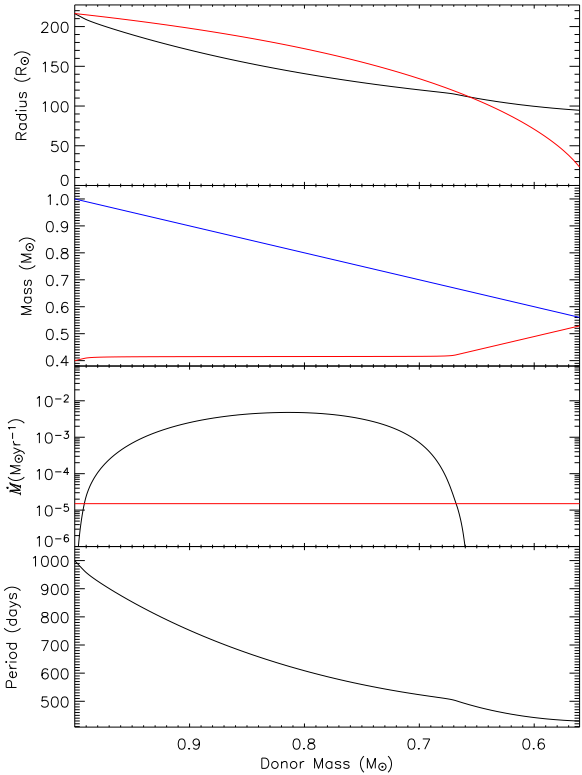,width=8.4cm}                                                    
\caption{Figure showing the evolution of a 1\,M$_{\sun}$ donor and a                    
  0.4\,M$_{\sun}$ white dwarf with an initial orbit period of                           
  1000\,days. The top panel shows the evolution of the radius of the                    
  donor (red line) and the donor's Roche-lobe (black) with donor                        
  mass. The second panel shows the masses of the components (blue and                   
  red for donor and accretor respectively). The third panel shows the                   
  mass transfer rate and the Eddington limit (red line). The bottom                     
  panel shows the evolution of the orbital period.} \label{figwd1000}                   
\end{center}                                                                            
\end{figure}                                                                            
                                                                                        
\begin{figure}                                                                          
\begin{center}                                                                          
\epsfig{file=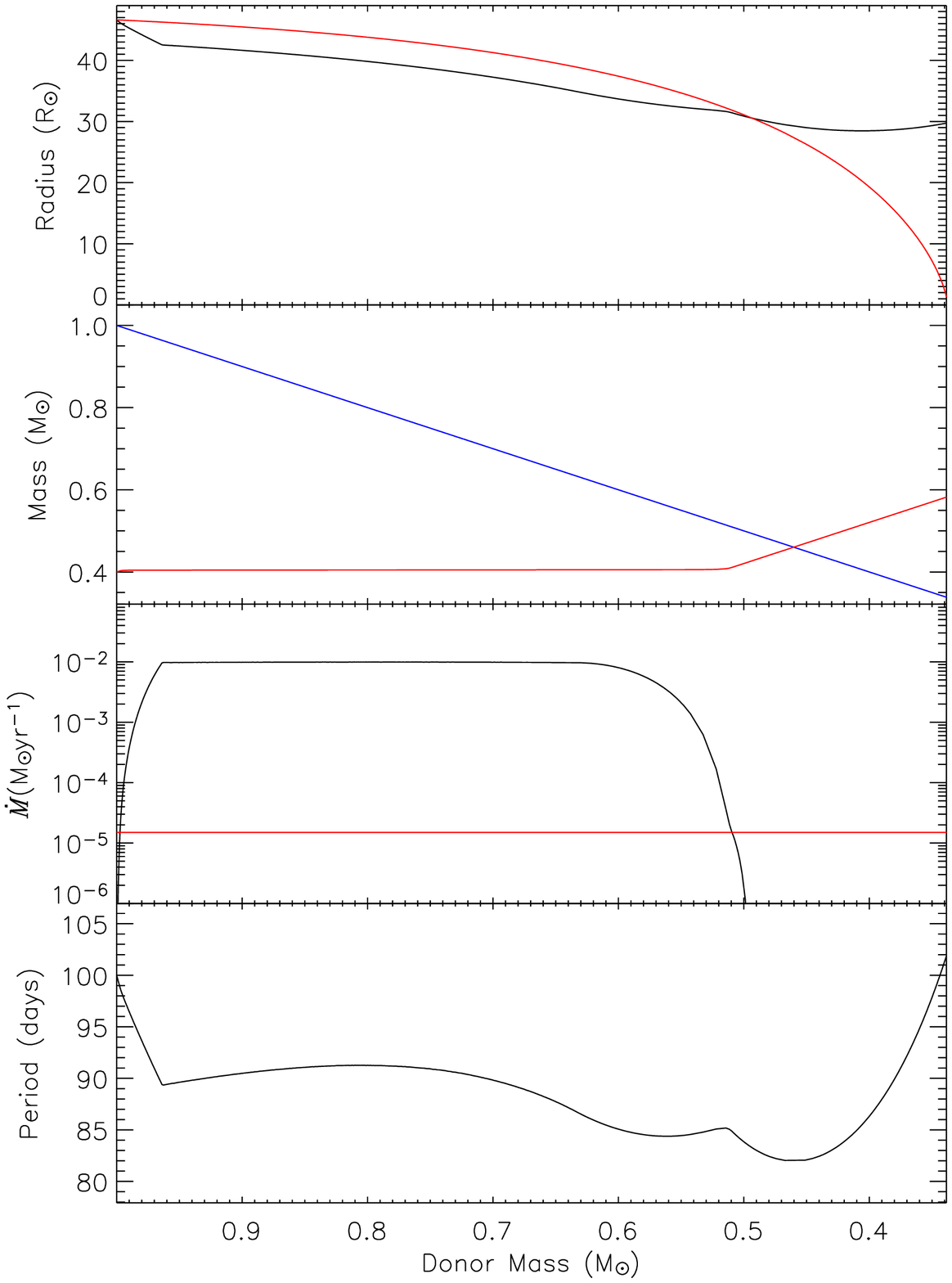,width=8.4cm}                                                     
\caption{Figure showing the evolution of a 1\,M$_{\sun}$ donor and a                    
  0.4\,M$_{\sun}$ white dwarf with an initial orbit period of                           
  100\,days. The nomenclature is the same as Figure~\ref{figwd1000}.}                   
\label{figwd100}                                                                        
\end{center}                                                                            
\end{figure}   

Once the donor fills its Roche lobe we consider the transfer of small portions
of the envelope $\Delta M$. The change in orbital separation of the system
$\Delta a$ is
\begin{equation}
  \frac{\Delta a}{a} = 2\frac{\Delta j}{j} + \frac{\Delta (M_1 +M_2) }
     {M_1+M_2} -2 \frac{\Delta M_1}{M_1} -2 \frac{\Delta M_2}{M_2} ~,
     \label{eqda} 
\end{equation}
where 
\begin{equation}
  \frac{\Delta j}{j} = -\frac{\psi q \Delta M \gamma} {(M_1 + M_2)} ~,
\end{equation}
and
\begin{eqnarray}
  \gamma = 
  \left \{
  \begin{array}{ll}
    1 & {\rm if } ~j = j_{\rm acc} \\
    \left [ 1 -
       \frac{\hbox{\rule[-3pt]{0pt}{12pt}$b$}}
     {\hbox{\rule[-3pt]{0pt}{12pt}$a$}} \left (
     \frac{\hbox{\rule[-3pt]{0pt}{12pt}$1+q$}}
     {\hbox{\rule[-3pt]{0pt}{12pt}$q$}} \right ) \right ] ^ 2 & {\rm
     if } ~j = j_{\rm L1}
  \end{array} \right .~, 
\end{eqnarray}
and equation~(\ref{eqda}) becomes
\begin{eqnarray}
  \frac{\Delta a}{a} = -\frac{2 \psi q \Delta M \gamma} {(M_1 + M_2)}
  - \frac{\psi \Delta M}{M_1+M_2} + \frac{2\Delta M} {M_1}
  \rule{40pt}{0pt} \nonumber\\
  -\frac{2(1-\psi) \Delta M}{M_2} ~.
\end{eqnarray}
We choose $\Delta M$ so that $\Delta a$ is always less than 1 per cent
of $a$. This formulation gives the same results as the calculations of
\nocite{bh91}{Bhattacharya} \& {van den Heuvel} (1991) and
\nocite{kw96}{Kalogera} \&  {Webbink} (1996) in which the final orbital
separation ($a_{\rm f}$) (after transfer of mass with a part $\psi$
lost with the angular momentum of the accretor) is given by
\begin{eqnarray}
  \frac{a_{\rm f}}{a_{\rm i}} = \left [
  \frac{\hbox{\rule[-3pt]{0pt}{12pt}$M_{2\rm{i}}$}}
  {\hbox{\rule[-2pt]{0pt}{10pt}$M_{2\rm{i}} + (1-\psi)
  (M_{1\rm{i}}-M_{1\rm{f}})$}} \right ] ^{\left (
  \frac{\hbox{\rule[-2pt]{0pt}{5pt}\scriptsize{$2$}}}
  {\hbox{\rule[0pt]{0pt}{5pt}\scriptsize{$1-\psi$}}}\right ) }\times
  \rule{40pt}{0pt} \nonumber\\ \left [ \frac
  {\hbox{\rule[-3pt]{0pt}{12pt}$M_{1\rm{i}} + M_{2\rm{i}}$}}
  {\hbox{\rule[-3pt]{0pt}{12pt}$M_{1\rm{i}} + M_{2\rm{i}} - \psi
  (M_{1\rm{i}}-M_{1\rm{f}})$}} \right ] \left ( \frac
  {\hbox{\rule[-3pt]{0pt}{12pt}$M_{1\rm{i}}$}}
  {\hbox{\rule[-3pt]{0pt}{12pt}$M_{1\rm{f}}$}} \right ) ^ 2 ~,
\end{eqnarray}
or by
\begin{eqnarray}
  \frac{a_{\rm f}}{a_{\rm i}} =
       \frac{\hbox{\rule[-3pt]{0pt}{12pt}$M_{1\rm{i}}+M_{2\rm{i}}$}}
       {\hbox{\rule[-3pt]{0pt}{12pt}$M_{1\rm{f}} + M_{2\rm{i}}$}} \left
       (\frac{\hbox{\rule[-3pt]{0pt}{12pt}$M_{1\rm{i}}$}}
       {\hbox{\rule[-3pt]{0pt}{12pt}$M_{1\rm{f}}$}} \right )^2 \exp\left [
       \frac{\hbox{\rule[-3pt]{0pt}{12pt}$-2 (M_{1\rm{i}}-M_{1\rm{f}})$}}
       {\hbox{\rule[-3pt]{0pt}{12pt}$M_{2\rm{i}}$}} \right ] ~,
       \label{eqbetaone}
\end{eqnarray}
if $\psi$ is one. The subscript i indicates the initial value before
mass transfer and $M_{1\rm{f}}$ is the final donor mass assumed to be
donor core mass upon filling its Roche lobe. The equations above are
identical with equations (A.5) and (A.7) of
\nocite{bh91}{Bhattacharya} \& {van den Heuvel} (1991) with the
substitutions $\psi \rightarrow \beta$ and appropriate notation for
the initial and final masses (e.g. $M_{1i} \rightarrow m_1^0$).

\begin{figure}                                                                       
\begin{center}                                                                       
\epsfig{file=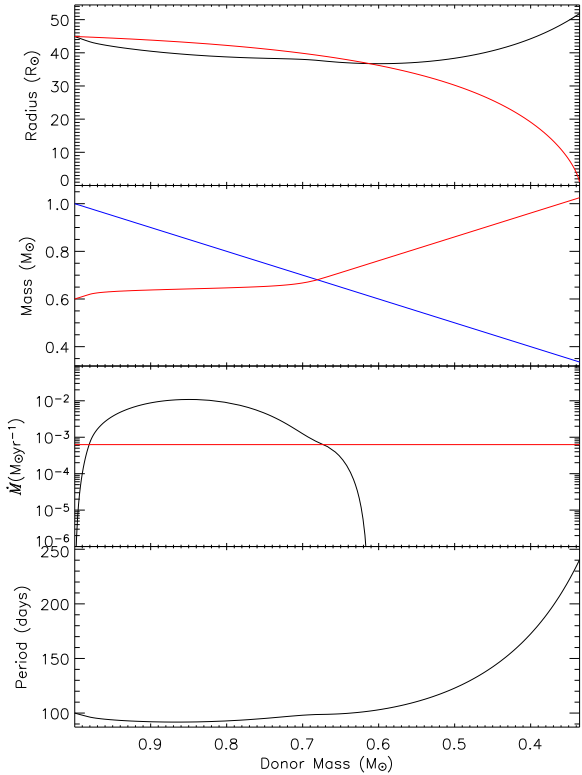,width=8.4cm}                                              
\caption{Figure showing the evolution of a 1\,M$_{\sun}$ donor and a      
  0.6\,M$_{\sun}$ accretor with an initial orbit period of                         
  100\,days. The nomenclature is the same as Figure~\ref{figwd1000}.}     
\label{figms100}                                                       
\end{center}                                
\end{figure}                                                    
                                                                                     
\begin{figure}                                             
\begin{center}                                          
\epsfig{file=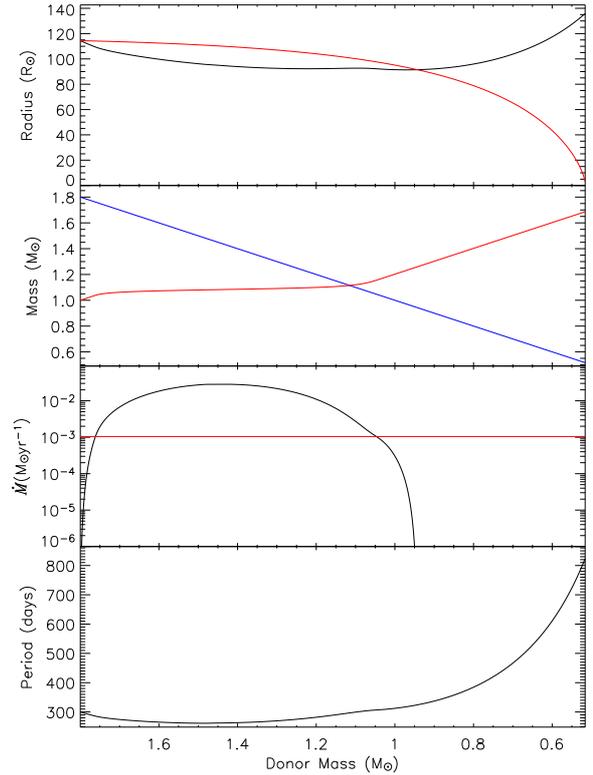,width=8.4cm}                                   
\caption{Figure showing the evolution of a 1.8\,M$_{\sun}$ donor and a 
  1.0\,M$_{\sun}$ accretor with an initial orbit period of             
  300\,days. The nomenclature is the same as Figure~\ref{figwd1000}.}
\label{figms300}                                        
\end{center}                                                      
\end{figure}   

After each change in orbital separation we calculate the new donor
radius, the corresponding new mass transfer rate
(equation~\ref{eqmdot}) and $\gamma$. The donor radius is found by
using the radius-mass exponents of the donor. \nocite{hw87}{Hjellming}
\& {Webbink} (1987) calculated radius-mass exponents $\zeta$ for
polytropic donor envelopes losing mass
rapidly. \nocite{sph97}{Soberman}, {Phinney} \& {van den Heuvel}
(1997) showed that these can be replaced by functions of fractional
core masses ($m$). We use the following prescription which is accurate
to better than a percent
\begin{eqnarray}
  \zeta = \frac{2}{3} \left ( \frac{m}{1-m} \right ) -
    \frac{1}{3} \left ( \frac{1-m} {1+2m} \right ) - 0.03 m \nonumber
    \rule{50pt}{0pt}\\
    + ~0.2 \left [ \frac{m}{1+(1-m)^{-6}} \right ] ~.
\end{eqnarray}
While condensed polytropic models \nocite{h89,h90}({Hjellming} 1989,
1990) generally represent the response of red giants well, there are
some cases where numerical models predict expansion. The resulting
larger transfer rate and trapping radius would increase the chance of
entering CE evolution. Evidently one would need to check for this in
individual cases.


\section{Results}
We consider the evolution of  1\,M$_{\sun}$ and 1.8\,M$_{\sun}$
donors transferring matter to four possible accretors. We choose these
donors because they have degenerate cores on the giant
branch. The accretors are (a) a low-mass main-sequence 
star with an initial mass of 0.2\,M$_{\sun}$ and a radius of
0.2\,R$_{\sun}$, (b) a white dwarf of mass 0.4\,M$_{\sun}$
and a radius of 10000\,km, (c) a main-sequence star of 0.6\,M$_{\sun}$
with an a radius of 0.6\,R$_{\sun}$, and finally a solar--type star
with initial mass 1\,M$_{\sun}$ and radius 1\,R$_{\sun}$.

Figures~\ref{figperiodinit} and~\ref{figperiodinit18} show the periods
at which the donors fill their Roche lobes and their corresponding core
masses. Black circles indicate the low-mass main-sequence companion,
red stars the white dwarf, green triangles the 0.6\,M$_{\sun}$
main-sequence star and blue squares the Sun-like star.

Figure~\ref{figseq100yr} shows the outcomes when mass is lost from a
1\,M$_{\sun}$ donor. Most low-mass main-sequence stars end up with 
the trapping radius greater than their Roche-lobe radii and so enter
CE evolution, producing a merged object with a portion of the donor
envelope ejected from the system before the CE phase. For the
white dwarf companions around 0.2\,M$_{\sun}$ is accreted during
the entire evolution and the majority of this after the dynamical
mass-transfer phase (see below). The more massive main-sequence stars
accrete significant mass. This mass is accreted, however, after the
dynamical mass-transfer such that the stars can cool the material
sufficiently as they accrete it. In a number of these evolutions the
mass-ratio is inverted during the dynamical mass-transfer phase which
brings an abrupt end to the dynamical phase. The systems then evolve
to longer periods (see below). 

Figure~\ref{figseq100yr1.8} shows the outcomes when mass is lost from
a 1.8\,M$_{\sun}$ donor. Only some of the white-dwarf and 
main-sequence companions survive. The white-dwarf accretors gain
little mass while the main-sequence accretors gain significantly at
the end of the dynamical mass-transfer phase (see below). We now
consider the evolution of specific systems which illustrate the
general trends shown in these sequences.

Figure~\ref{figwd1000} shows the evolution of a white-dwarf accretor
with an initial orbital period of 1000 days. The donor overfills its
Roche lobe on the asymptotic giant branch (AGB) with a core mass of
0.56\,M$_{\sun}$. After transferring a small portion of mass on the
nuclear expansion timescale for the giant the donor becomes
sufficiently oversized with respect to its Roche lobe that dynamical
mass transfer occurs. During this phase 0.3\,M$_{\sun}$ of the
envelope is transferred reaching peak mass transfer rates of just under
10$^{-2}$\,M$_{\sun}$yr$^{-1}$. The white dwarf accretes at the
Eddington rate during this phase and as it is a rapid phase (a
timescale of order 100\,yrs) the white dwarf does not gain much mass
compared to its initial mass. After this dynamical phase however, the star has
contracted back inside its Roche lobe and mass transfer is limited to
the nuclear expansion of the envelope again. We consider this portion
of the evolution to be conservative (all transferred mass is accreted)
and the remaining 0.1\,M$_{\sun}$ of the envelope is accreted on to the
white-dwarf. 

Figure~\ref{figwd100} shows the evolution of a white-dwarf accretor
with an initial orbital period of 100 days. The donor overfills its
Roche lobe on the giant branch with a core mass of
0.35\,M$_{\sun}$. During the dynamical mass-transfer phase
0.45\,M$_{\sun}$ of the 
envelope is transferred reaching peak mass transfer rates of 
10$^{-2}$\,M$_{\sun}$yr$^{-1}$. At this mass-transfer rate the
trapping radius (which only depends on the transfer rate) fills the
circularization radius of the accretor. The system then switches to
losing mass with the specific angular momentum of the inner Lagrangian
point which tends to widen the binary (see bottom panel of
Figure~\ref{figwd100}). The evolution through this phase then
continues so that $R_{\rm trap}$ is the same as $R_{\rm circ}$. If the
trapping radius is larger than the circularization radius then the
binary widens and the transfer rate drops while if is smaller the
binary tightens and the transfer rate increases. After this dynamical
phase the remaining 0.18\,M$_{\sun}$ of the envelope is accreted on to
the white-dwarf. As the mass ratio is inverted the binary becomes
wider.

Once the dynamical mass-transfer phase has finished the evolution of
the system will proceed much more slowly. Clearly the assumption that
the core mass stays constant is no longer valid. The only effect this
will have on the calculations, however, is that less mass is available
to transfer. This will result in evolution ending sooner and the
systems not widening as much after the dynamical mass-transfer
stage. In the case of white-dwarf accretors the mass transfer has been
assumed to be conservative. During the dynamical mass-transfer phase
the accretion rates will be much higher than those required for stable
nuclear burning \nocite{kh97}(see {Kahabka} \& {van den Heuvel} 1997)
but afterwards may be low enough for nova behaviour. In this case the
final part of the mass-transfer may be non-conservative and the
systems would end up tighter i.e. shorter periods.

Figure~\ref{figms100} shows the evolution of a 0.6\,M$_{\sun}$
main-sequence star with an initial orbital period of 100 days. The
donor overfills its 
Roche lobe on the giant branch with a core mass of
0.34\,M$_{\sun}$. During the dynamical mass-transfer phase
0.3\,M$_{\sun}$ of the 
envelope is transferred reaching peak mass transfer rates of under
10$^{-2}$\,M$_{\sun}$yr$^{-1}$. Near the end of this dynamical phase
however, the mass ratio has been inverted and mass transfer widens the
binary rapidly halting the dynamical mass-transfer phase. The
remaining 0.35\,M$_{\sun}$ of the envelope is then accreted on to the 
main-sequence accretor raising its mass to over 1\,M$_{\sun}$. The
evolution of this system is substantially different to that of
standard CE evolution where the accretor would be assumed not to gain
in mass. 

\begin{figure*}                                                         
\begin{center}                                                          
\subfigure{\epsfig{file=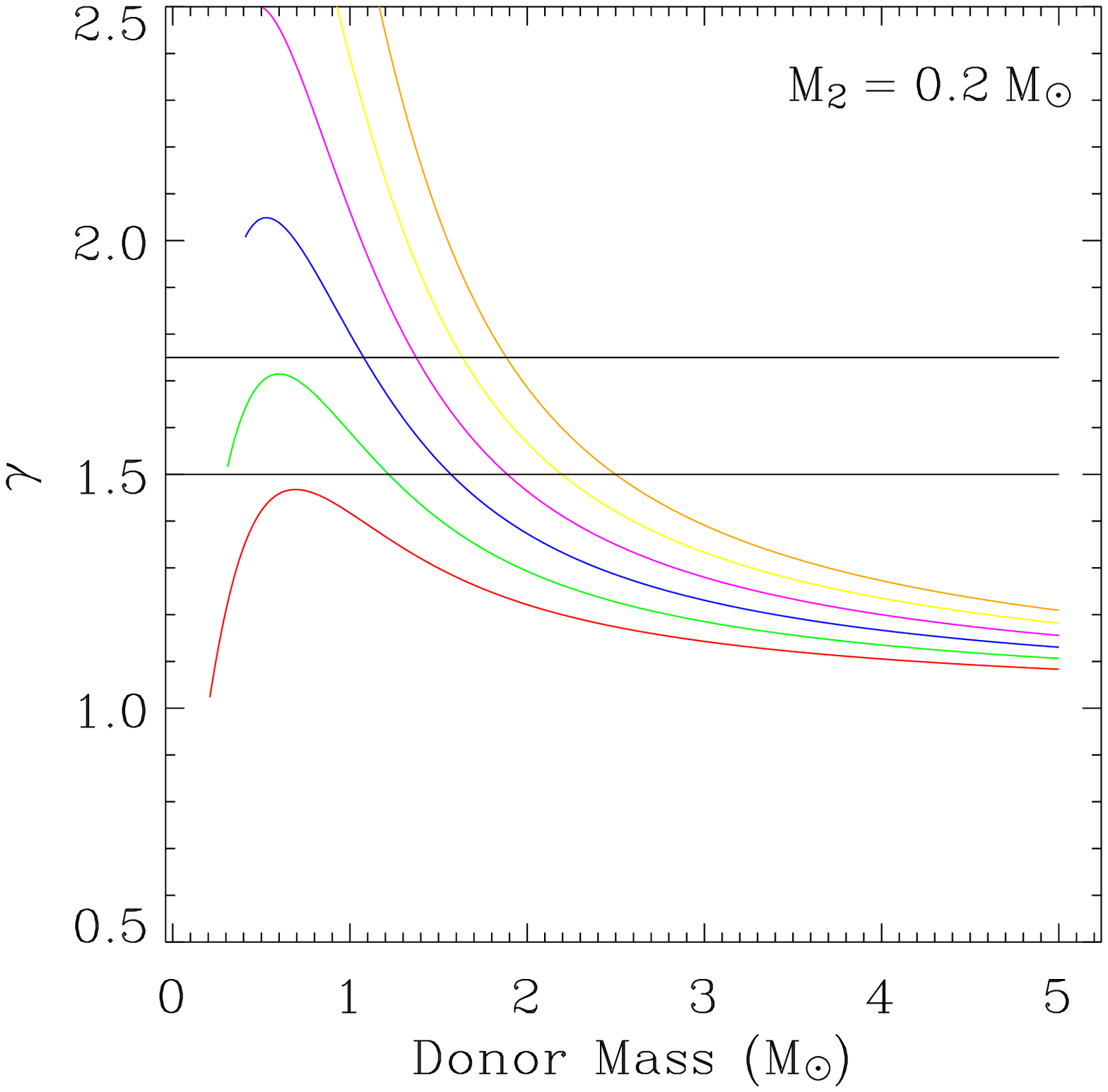,height=5.5cm,}\label{fig0.2gamma}} 
\subfigure{\epsfig{file=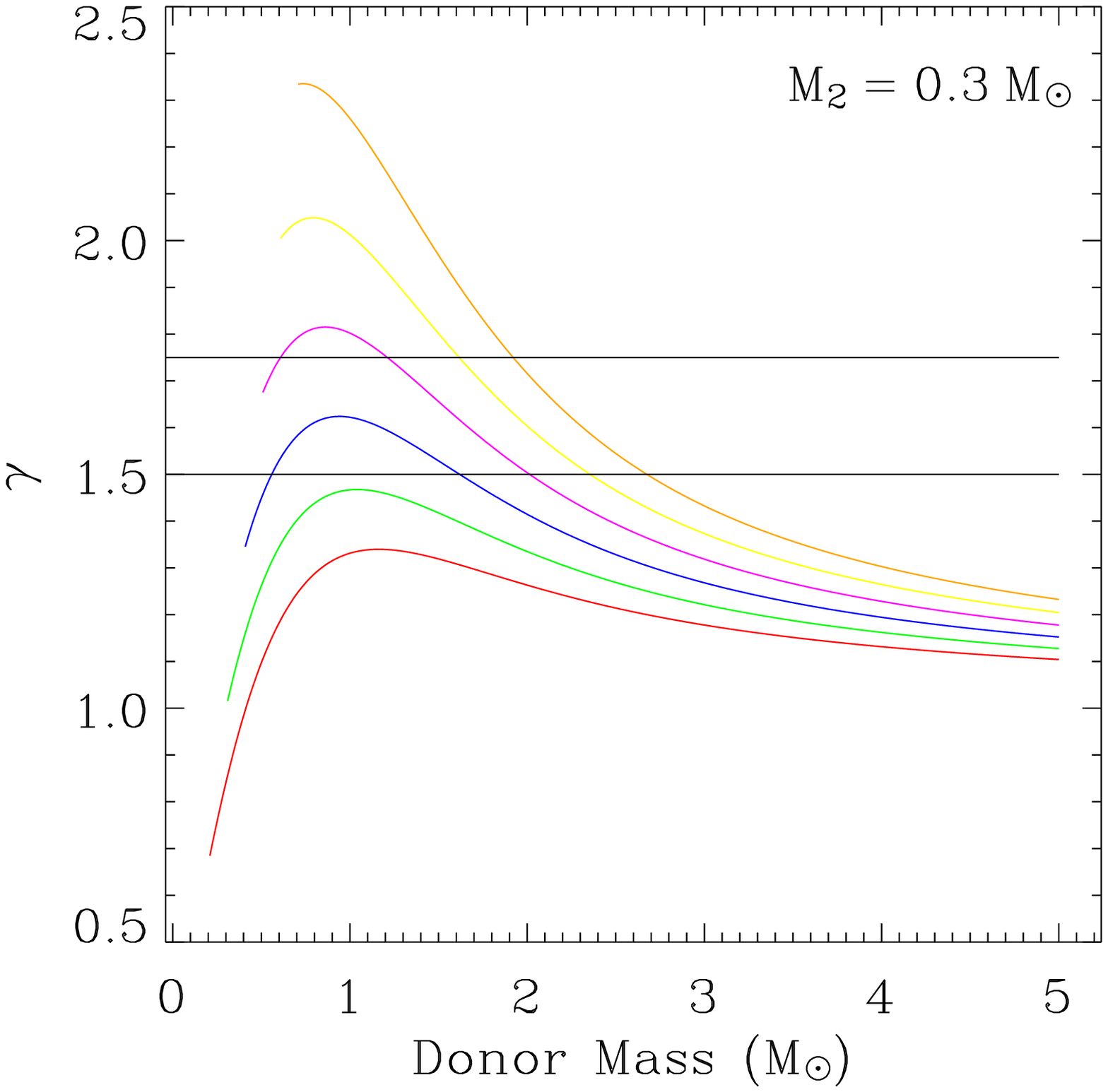,height=5.5cm,}\label{fig0.3gamma}} 
\subfigure{\epsfig{file=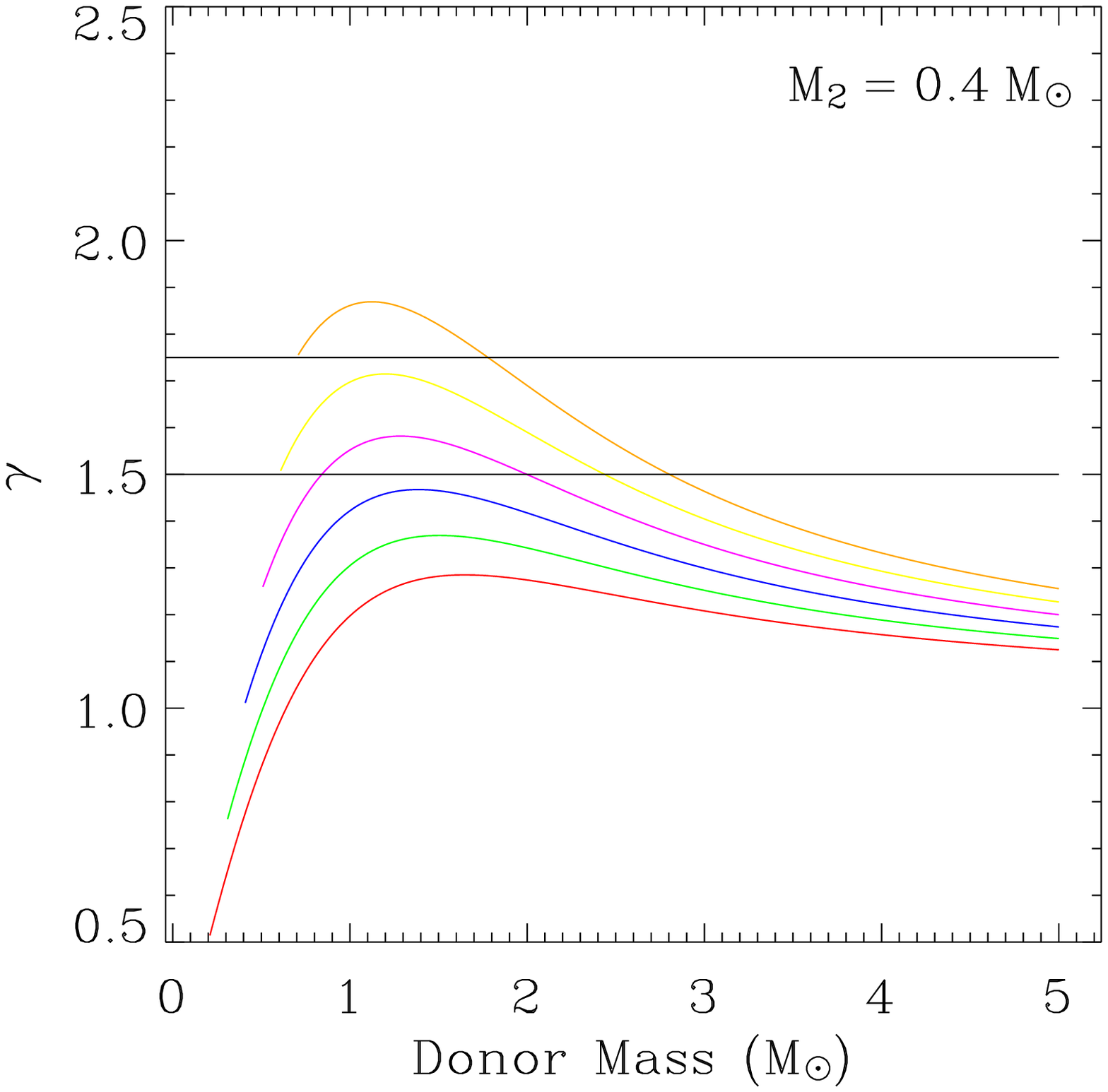,height=5.5cm,}\label{fig0.4gamma}} 
\subfigure{\epsfig{file=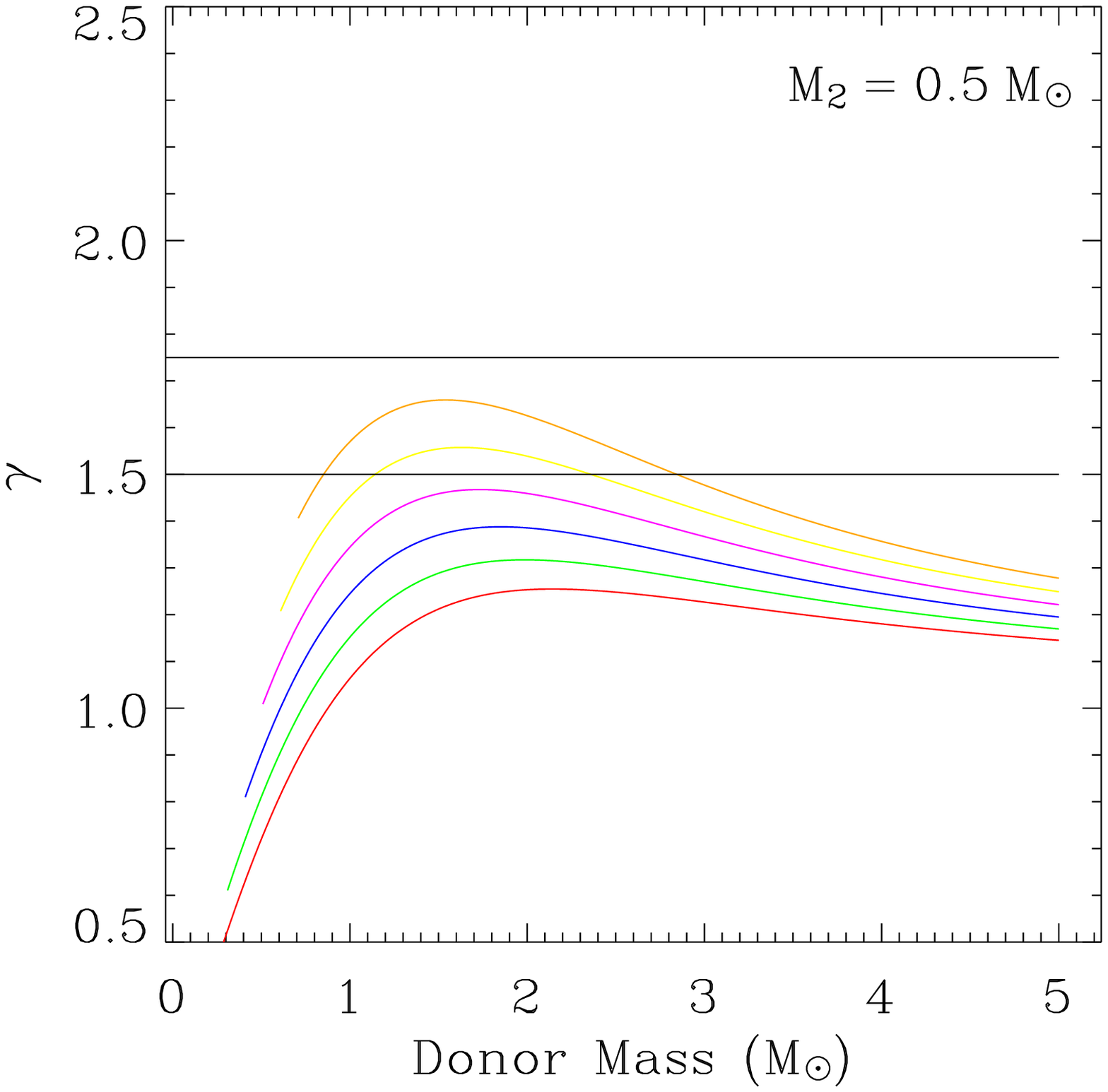,height=5.5cm,}\label{fig0.5gamma}} 
\subfigure{\epsfig{file=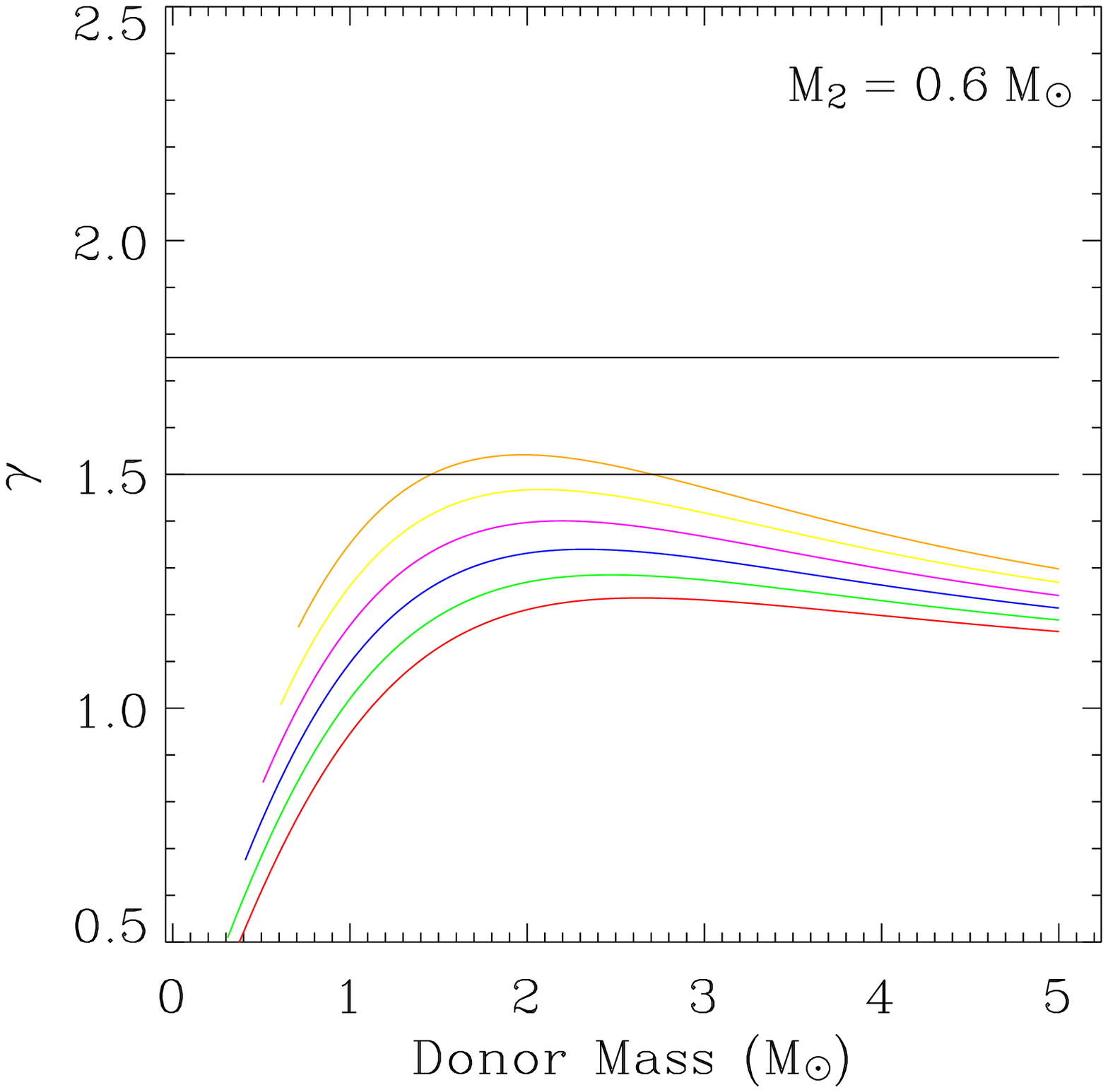,height=5.5cm,}\label{fig0.6gamma}} 
\caption[Values of $\gamma_{\rm max}$ for accretors ranging from 0.2    
  to 0.6\,M$_{\sun}$. All the transferred matter is assumed lost from   
  the system carrying the angular momentum of the accretor. Line
  colour\
s                                                                       
  indicate donor core masses of 0.2 (bottom) to 0.7\,M$_{\sun}$ (top)   
  in increments of 0.1\,M$_{\sun}$.Solid                                
  horizontal lines show the range for $\gamma$ suggested by]{Values of  
  $\gamma_{\rm max}$ for                                                
  accretors ranging from 0.2 to 0.6\,M$_{\sun}$. All matter lost from   
  the system is assumed to carry the angular momentum of the            
  accretor. Line colours indicate donor core masses of 0.2 (bottom) to  
  0.7\,M$_{\sun}$ (top) in increments of 0.1\,M$_{\sun}$. Solid         
  horizontal lines show the range for $\gamma$ suggested by             
  \nocite{nt05}{Nelemans} \&  {Tout} (2005).} \label{figgamma}          
\end{center}                                                            
\end{figure*}    

Figure~\ref{figms300} shows the evolution of a binary composed of a
1.8 and a 1.0\,M$_{\sun}$ main-sequence star with an initial orbital
period of 300 days. As in Figure~\ref{figms100}, near the end of
the dynamical phase the mass ratio is inverted. The system then
evolves to longer periods. 

The radial response of low-mass main-sequence star to accretion is
significantly different to that of a high-mass star. As
\nocite{fi89}{Fujimoto} \& {Iben} (1989) show a low-mass star with a
convective envelope does not significantly expand on accretion and
could even contract. In our scenario the entropy of the transferred
mass will be higher than the main-sequence surface but it will be
subject to a far higher surface gravity. Regardless of the detailed
radial response of the secondary it is clear from
Figures~\ref{figms100} and \ref{figms300} that only a relatively small
portion of mass is accreted in the build-up to and during dynamical
mass-transfer and so we do not expect significantly expand during this
process.

\subsection{Parametrizations}

Treatments of CE evolution have generally been expressed in terms of
empirical parametrizations \nocite{w84,nt05}({Webbink} 1984;
{Nelemans} \& {Tout} 2005). Here we compare our results with these
parametrizations. This serves two purposes. First, where the
parametrizations are successful, our treatment should agree, and
indeed offer an explanation for the success. Second, a simple and
robust parametrization would give a succinct way of summarizing
complex physics. As these parametrizations assume no accretion onto
the secondary we make the same assumption in order to compare our
model to their parametrizations.

\begin{figure*}                                                                         
\begin{center}                                                                          
\subfigure{\epsfig{file=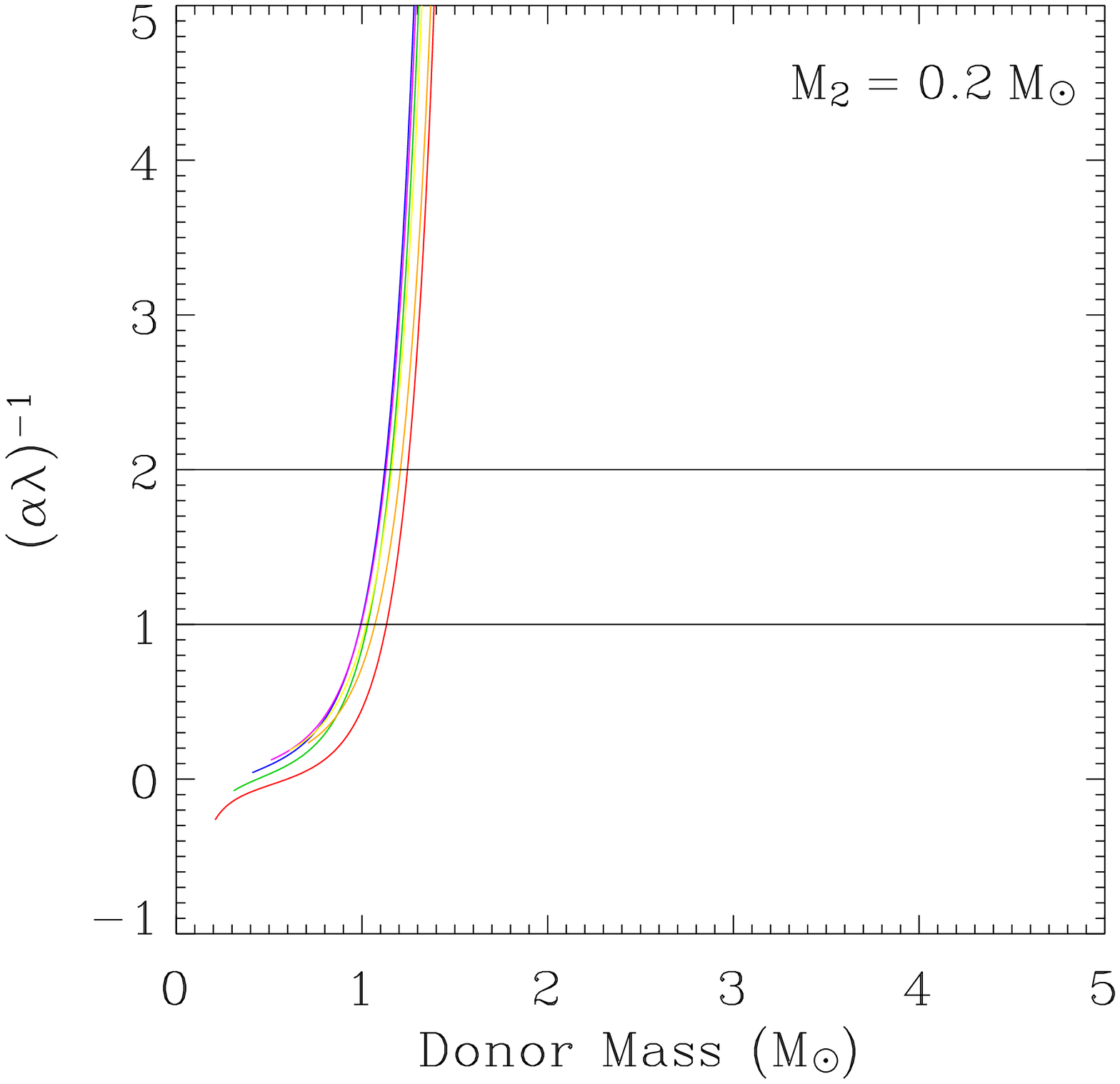,height=5.5cm,}\label{fig0.2alpha_lambda}}   
\subfigure{\epsfig{file=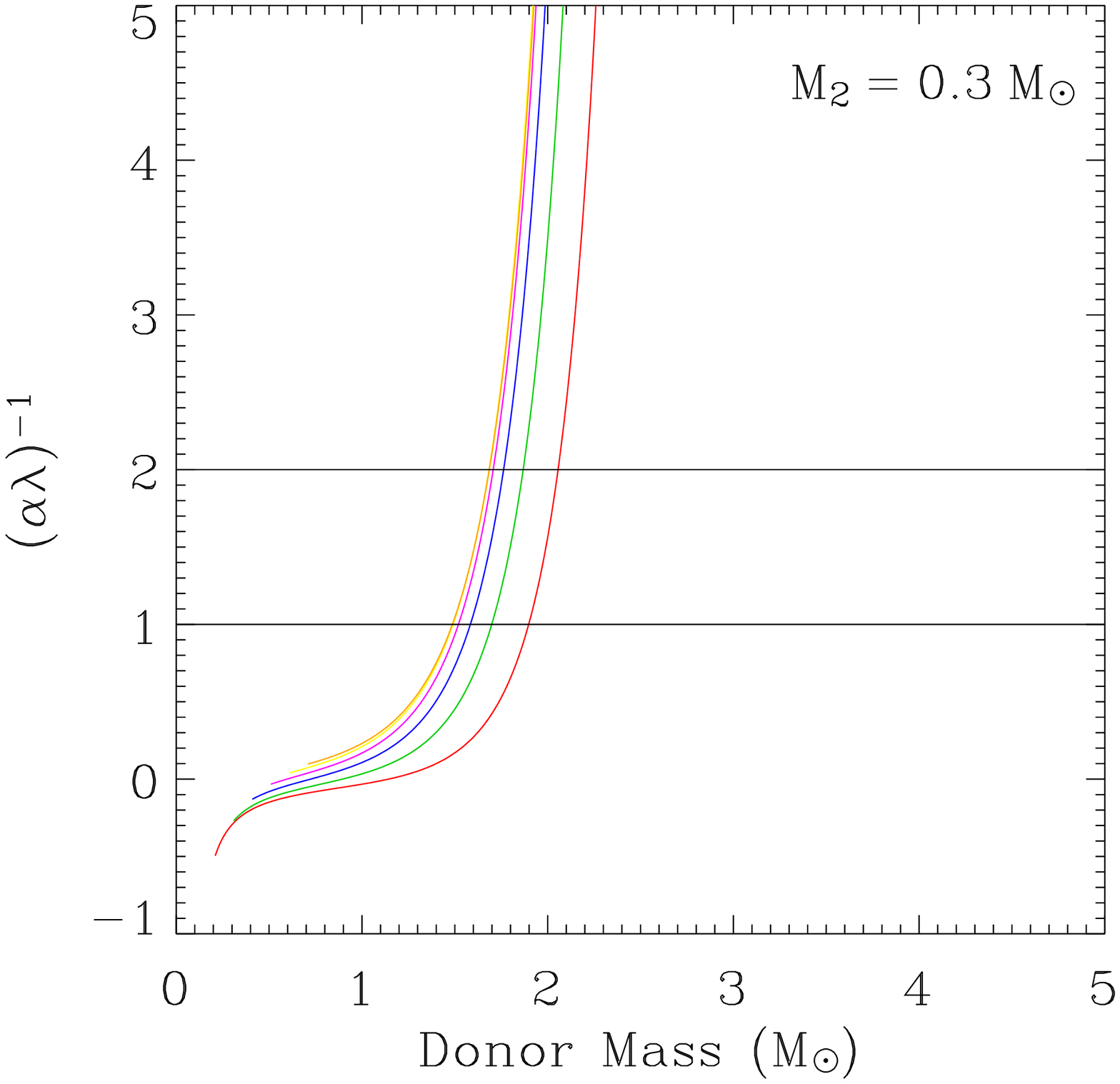,height=5.5cm,}\label{fig0.3alpha_lambda}}   
\subfigure{\epsfig{file=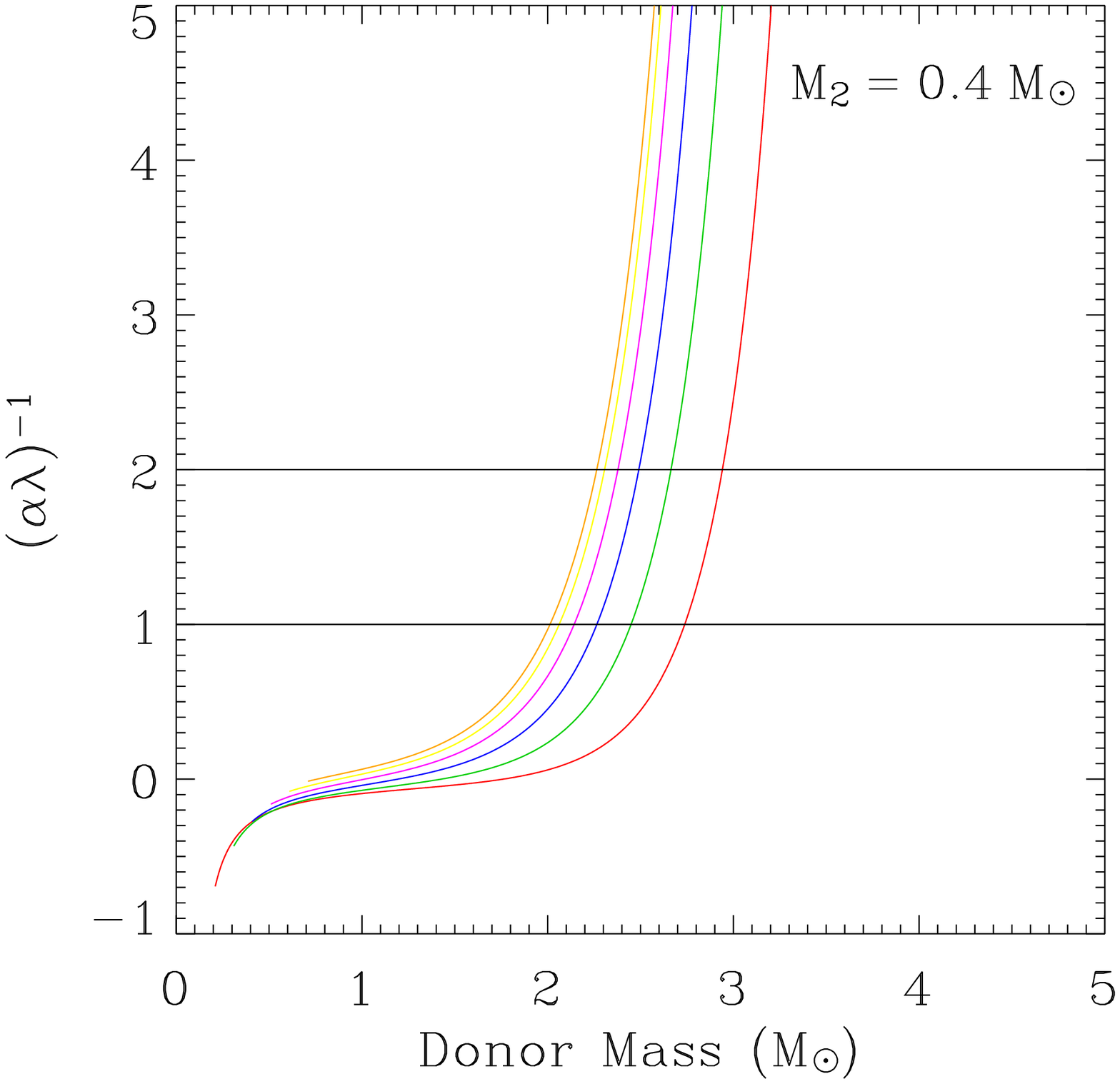,height=5.5cm,}\label{fig0.4alpha_lambda}}   
\subfigure{\epsfig{file=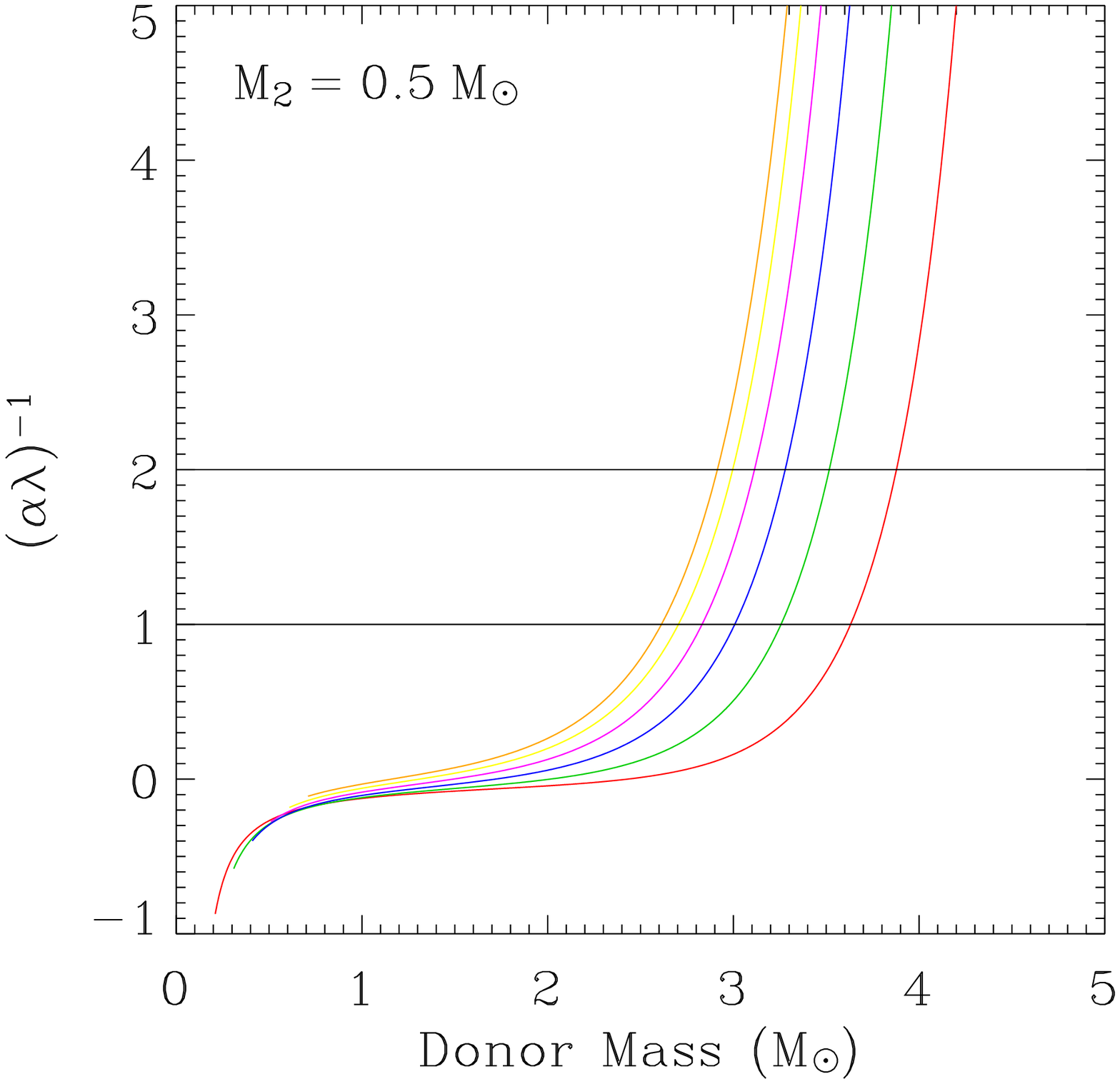,height=5.5cm,}\label{fig0.5alpha_lambda}}   
\subfigure{\epsfig{file=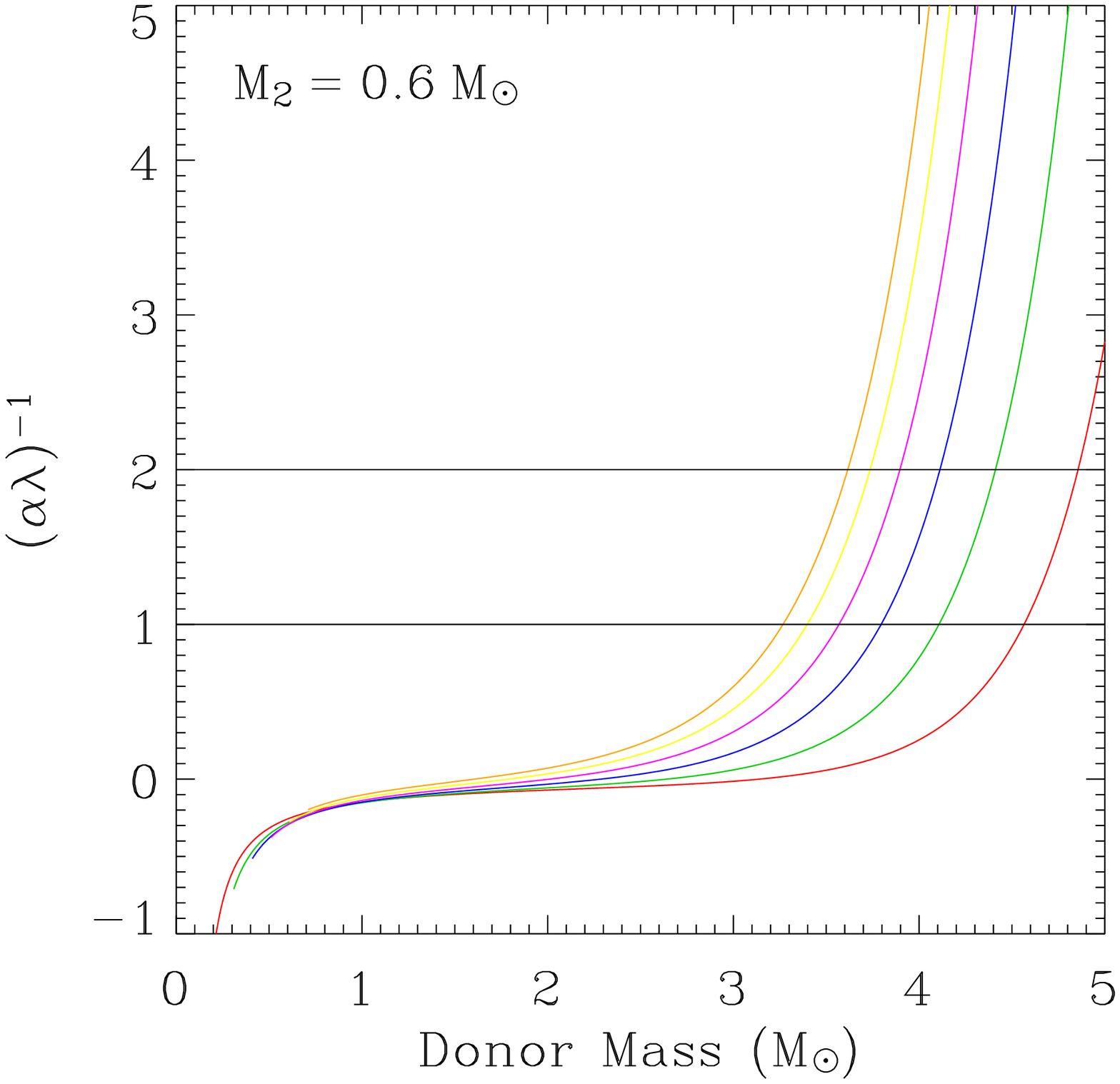,height=5.5cm,}\label{fig0.6alpha_lambda}}   
\caption{Values of $(\alpha\lambda)^{-1}$ for accretors ranging from                    
  0.2                                                                                   
  to 0.6\,M$_{\sun}$. All the transferred matter is assumed lost from                   
  the system carrying the specific angular momentum of the                              
  accretor. Line colours                                                                
  indicate donor core masses of 0.2 (bottom) to 0.7\,M$_{\sun}$ (top)                   
  in increments of 0.1\,M$_{\sun}$. Solid                                               
  horizontal lines show the typical values assumed for $\alpha\lambda$                  
  of 0.5--1.0.}\label{figalphalambda}                                                   
\end{center}                                                                            
\end{figure*} 

Our picture predicts a total change in angular momentum ($\Delta J$)
proportional to the total change in system mass ($\Delta M$)
\begin{equation}
\frac{\Delta J}{J} = \gamma \frac{\Delta M}{M} \,. \label{eqgamma}
\end{equation}
This is the form assumed empirically by
\nocite{nt05}{Nelemans} \&  {Tout} (2005). Our
physical 
picture of the process allows us to 
specify an upper limit on $\gamma$ which we can compare with the
results they get by applying their parametrization to observed
systems. 

Assuming that the entire envelope of the donor is lost (\ref{eqgamma}) becomes
\begin{equation}
\frac{\Delta J}{J} = \gamma \frac{M_{\rm env}} {M_{\rm c} + M_2} \, .
\end{equation}
The orbital angular momentum of the system is 
\begin{equation}
J = M_1 M_2 \sqrt {\left . \frac{G a} {M_1 + M_2} \right . } \, ,
\end{equation}
so the left hand side of equation~(\ref{eqgamma}) is
\begin{equation}
\frac{\Delta J}{J_{\rm i}} = 1 - \frac{M_{1\rm f}}{M_{1\rm i}}
\frac{M_{2\rm f}}{M_{2\rm i}} \sqrt{\left . \frac{ ( M_{1{\rm i}} +
    M_{2{\rm i}}  ) a_{\rm f} } { ( M_{1{\rm f}} +
    M_{2{\rm f}}  ) a_{\rm i} } \right . } \,.
\end{equation}
If the matter is lost with the angular momentum of the
  accretor\footnote{\nocite{s06}{van der Sluys}, {Verbunt} \& {Pols}
  (2006) have applied the method of \nocite{nt05}{Nelemans} \& {Tout}
  (2005) to a larger number of systems and find the observed double
  white-dwarfs can be well explained by assuming that envelope
  ejection occurs with the specific angular momentum of the accretor.}
  then taking $M_2$ as fixed, and using equation~(\ref{eqbetaone})
  this becomes
\begin{equation}
\frac{\Delta J}{J_{\rm i}} = 1 - \frac{M_{1\rm i} + M_2}
     {M_{1\rm f} + M_2} \exp {\left ( - \frac{M_{\rm{env}}} {M_2} 
     \right ) } \,. 
\end{equation}
Hence in this case we have
\begin{equation}
\gamma = \frac{M_{1\rm i} + M_2} {M_{\rm env}} - \frac {
  \left ( M_{1\rm i} + M_2 \right ) ^2 } {M_{\rm env} (M_{\rm c} + M_2) } \exp
  { \left ( - \frac{M_{\rm{env}}} {M_2} 
     \right ) } \,. \label{eqgammaeff}
\end{equation}
This is an upper limit as we assume that the lost matter carries the
specific angular momentum of the accretor which will have a higher
specific angular momentum than the inner Lagrangian point. Hence if
there is any mass loss without a disc the specific angular momentum of 
the lost matter will be smaller and the total $\Delta J$ will be less.

Fig.~\ref{figgamma} shows $\gamma$ versus donor mass for various core
and accretor masses. \nocite{nt05}{Nelemans} \& {Tout} (2005) show
that observed close white--dwarf binaries are well described by the
choice $\gamma \simeq$ 1.5\,--\,1.75. Since many of the values in
Fig.~\ref{figgamma} are within this range we conclude that the
evolution of close white-dwarf binaries is compatible with the idea
that all the transferred matter is ejected with the specific angular
momentum of the accretor.

\nocite{nt05}{Nelemans} \&  {Tout} (2005) considered neutron--star
 progenitors in the antecedents of low--mass X--ray binaries and
 argued that $\gamma$ values slightly above 1.5 allow formation. From
 equation~(\ref{eqgammaeff}) we find that a typical system with
 $M_{\rm env} = 6$\,M$_{\sun}$, $M_{\rm c} = 2.5$\,M$_{\sun}$, $M_2$ =
 1\,M$_{\sun}$ has $\gamma = 1.56$. This is low enough not
 to merge, but high enough to form a tight binary. Again these systems
 are compatible with all the transferred matter being ejected with the
 angular momentum of the accretor.

We have not considered accretion on to neutron stars but such accretion
is likely to be super--Eddington so we would expect most of the mass
to be ejected with the specific angular momentum of the neutron
star. This evolution is similar to Cyg X--2. Applying the results of
\nocite{kr99}{King} \&  {Ritter} (1999) with $M_{1i} = 3.6\,{\rm
M_{\sun}},~M_{\rm env} = 3\,{\rm M_{\sun}},~M_2 = 1.4\,{\rm
M_{\sun}}$, equation~(\ref{eqgammaeff}) gives a $\gamma$ of
1.2.

We now turn to the comparison with the energy or $\alpha\lambda$
formalism \nocite{w84}({Webbink} 1984).  Making the same assumptions as for
equation~(\ref{eqgammaeff}) (all mass ejected with specific angular
momentum of the accretor) we find 
\begin{eqnarray}
\frac{1}{\alpha\lambda} = \left \{ \frac{M_{\rm c}+M_2} {M_{\rm 1i}+M_2}
\left ( \frac{M_{\rm c}}{M_{\rm 1i}} \right )^2 \exp\left [ \frac {2
    \left (M_{\rm 1i} - M_{\rm c}\right ) } {M_2} \right ] - 
    \right . \rule{15pt}{0pt} \nonumber\\ 
      \left . \frac{M_{\rm 1i}} {M_{\rm c}} \right \} 
      \frac { M_{\rm c}  M_2
      \left ( R_{\rm 1i}/a_{\rm i} \right ) } {2 M_{\rm 1i} \left (
	M_{\rm 1i} - M_{\rm c} \right ) } ~,
\end{eqnarray}
where $R_{\rm 1i}/a_{\rm i}$ is given by equation~(\ref{eqeggrl}) and is
a function of only the initial mass ratio ($M_{\rm 1i}/M_2$).
Fig.~\ref{figalphalambda} shows the values of $(\alpha\lambda)^{-1}$
using these assumptions. The solid lines indicate the likely range of
$\alpha\lambda$ of 0.5\,--\,1.0. Again one sees that at least some
choices of $\alpha\lambda$ produce 
the same outcomes as predicted by our calculations. However, it is clear that
for different accretor masses only a small range of donor masses are
possible. This contrasts to the broad range of donor masses
possible with the $\gamma$ prescription. Clearly $\gamma$ is not a
constant but a free parameter which weakly constrains the possible
progenitors of a system compared to the $\alpha\lambda$
formalism. Our calculations may offer a 
physical justification for these parametrizations.  However it is
clear that neither parametrization gives a full description of our
results, which is perhaps not surprising. 

\section{Conclusions}

We have studied binary evolution in cases which are normally thought
to lead to common--envelope evolution. We have shown instead that the
accretion luminosity is in many cases able to expel much of the
super--Eddington mass transfer without catastrophic orbital shrinkage.
Treatments of CE evolution have hitherto relied on empirical 
parametrizations \nocite{w84,nt05}(e.g. {Webbink} 1984; {Nelemans} \&  {Tout} 2005). In cases 
where these are successful in describing observed systems our theoretical 
treatment agrees quite well with the adopted parametrizations. This 
suggests that our ideas give a reasonable description of the prior 
evolution of cataclysmic variables, double white dwarf binaries and 
low--mass X--ray binaries.

\section*{Acknowledgements}
MEB and LMD are PPARC research associates. ARK acknowledges a Royal
Society--Wolfson Research Merit Award. We thank the referees for
suggestions which considerably improved this paper.


{\vspace{0.5cm}\small\noindent This paper
has been typeset from a \TeX / \LaTeX\ file prepared by the author.}

\label{lastpage}

\end{document}